\documentclass[11pt,letterpaper]{article}
\usepackage[margin=1in]{geometry}
\usepackage[T1]{fontenc}
\usepackage{lmodern,amsmath,amssymb,booktabs,longtable,array,graphicx}

\usepackage[authoryear,round]{natbib}
\usepackage{microtype,placeins}
\usepackage[font=small,labelfont=bf]{caption}
\usepackage{xurl}
\usepackage[hidelinks]{hyperref}
\renewcommand{\arraystretch}{1.15}

\newcolumntype{P}[1]{>{\raggedright\arraybackslash}p{#1}}

\hypersetup{
 pdftitle={Candidate-dependent residual routing trades top-of-list accuracy for recall in sequential recommendation},
 pdfauthor={Han Chen and Yingrui Li},
 pdfsubject={Sequential recommendation; residual routing; ranking evaluation},
 pdfkeywords={sequential recommendation, multi-behavior recommendation, candidate-dependent routing, recall, ranking evaluation}
}
\title{Candidate-dependent residual routing trades top-of-list accuracy for recall in sequential recommendation}
\author{Han Chen$^{1,\dagger}$ and Yingrui Li$^{2,\dagger}$\\[0.3em]
\normalsize $^1$ Department of Computer Science,\\
\normalsize The University of Texas at Austin, Austin, Texas, United States\\
\normalsize $^2$ Unaffiliated, Fairfax, Virginia, United States\\
\normalsize $^\dagger$ These authors contributed equally.\\[0.3em]
\normalsize Corresponding author: Han Chen, \texttt{han.chen5@utexas.edu}}
\date{}
\begin{document}
\raggedright
\setlength{\parindent}{1.25em}
\maketitle
\begin{abstract}
Candidate-dependent routing lets a sequential recommender weight a user's past behaviors differently for each item it scores. We test whether this flexibility helps by adding a small residual routing module to a self-attentive baseline, in two versions that differ only in what sets the routing weights: the user's history representation or the candidate item's embedding. Across five paired seeds on the REES46 cosmetics-shop dataset, with matched training for all three models, candidate-dependent routing leaves the primary metric, normalized discounted cumulative gain at cutoff ten (NDCG@10), essentially unchanged (mean relative change +0.003\%). Secondary and exploratory analyses show that this flat average hides two effects that cancel. The candidate-dependent model places more relevant items in its top ten (recall +1.81\%) but ranks the first one lower (mean reciprocal rank $-1.09$\%), and most of its new hits land at positions six to ten. Its gains come from queries with at least one target among the user's last 50 events, while queries whose targets all lie outside that window get worse. In this form, candidate-dependent routing trades accuracy at the top of the list for recall further down.

\end{abstract}

\section{Introduction}

A shopper's history mixes several kinds of evidence. On a cosmetics site, the same lipstick can be viewed, added to a cart, bought or removed from the cart, and each action says something different about what the shopper will want next. A sequential recommender has to turn that history into a score for every candidate item. One design choice is whether to summarize the history once per user or to let each candidate decide which parts of the history matter for it.

Letting the candidate decide is appealing, because evidence that helps score one item may say little about another. Deep Interest Network weights past behaviors by their relevance to the candidate \citep{din}, and the Multi-Interest Network with Dynamic Routing (MIND) represents each user with several interest vectors \citep{mind}. But a self-attentive encoder already selects information from the history \citep{sasrec}. An extra candidate-dependent weighting step might duplicate what the encoder has learned, or reshuffle the ranking without improving the metric used to judge it.

We test this with a small residual routing module added to a baseline modeled on the Self-Attentive Sequential Recommendation model (SASRec). The module averages the item embeddings in each behavior stream (views; cart additions and purchases; cart removals), weights the streams with attention and adds the result to the baseline's score. The vector that sets the weights, which we call the routing vector, comes either from the user's history representation (user-dependent routing) or from the candidate's embedding (candidate-dependent routing). The two versions have the same number of parameters, so comparing them isolates the effect of letting the candidate choose, and comparing each with the baseline shows what routing adds at all.

On REES46, with five paired seeds and full-catalog evaluation, candidate-dependent routing barely changes our primary metric, normalized discounted cumulative gain at cutoff ten (NDCG@10): the average change is +0.003\%, and the one-sided 95\% upper bound of +0.49\% lies below the 1\% margin we set in advance. The average is flat because two effects, each consistent across seeds, cancel. The candidate-dependent model places more relevant items in its top ten but, on average, ranks the first one lower, so gains at ranks two to ten offset a loss at rank one. The gains also depend on the query: they come from queries with at least one target already in the user's recent history, while queries whose targets are all new to that history get worse. Candidate-dependent routing therefore changes what the model recommends, but not in a way that average NDCG@10 rewards.

We also revisit earlier Transformer models that apply candidate-dependent routing inside their attention blocks, on REES46, RetailRocket and KuaiSAR. These models use a different mechanism and were selected under different procedures, so we report them separately, as context for the main experiment rather than a second test of it.

\section{Related Work}

\subsection{Selecting information from interaction histories}

Sequential recommenders differ in how they keep and select past interactions. GRU4Rec processes a session with gated recurrent units \citep{gru}, whereas SASRec uses self-attention to pick out relevant earlier items \citep{sasrec}. Both summarize the history in one representation that is used to predict the next interaction. Our baseline follows the self-attentive design.

Candidate-dependent methods let the same history produce a different representation for each item being scored. Deep Interest Network weights past behaviors with a learned attention function, its local activation unit, according to their relevance to a candidate advertisement \citep{din}. MIND routes history embeddings into several interest vectors, and during training the target item determines how these vectors are weighted \citep{mind}. Our module is simpler: it averages the item embeddings in each predefined behavior stream, weights these averages and adds the result to an existing sequential score.

\subsection{Behavior types and repeated interactions}

Multi-behavior recommenders model relationships among actions as well as among items. The Multi-Behavior Hypergraph-Enhanced Transformer (MBHT) combines a multi-scale Transformer with a hypergraph to capture dependencies across behavior types \citep{mbht}, and the Multi-Behavior Sequential Transformer Recommender (MB-STR) pairs a multi-behavior Transformer with a sequential-pattern component and behavior-aware prediction \citep{mbstr}. Search-aware recommendation separates what can be learned from search histories and from recommendation histories \citep{sesrec}. Each of these designs changes the history representation, the training signal or both. We keep the encoder architecture and training procedure fixed and change only the source of the routing vector, which isolates one design choice rather than comparing whole architectures.

Repeat behavior is a second source of variation. RepeatNet explicitly chooses between recommending an item from the session history and exploring new items \citep{repeatnet}. Our models make no such explicit choice, but we examine whether their predictions differ for targets that do or do not appear in the recent history the model sees.

\subsection{Evaluating incremental improvements}

Whether a new component helps depends on the strength and tuning of the baseline and on the evaluation protocol \citep{progress,standards}. Metrics computed by ranking a sampled subset of items can disagree with full-catalog metrics \citep{sampling}, and the way ties are broken can change a result even when the scores do not \citep{ties}. We therefore rank the full catalog, break exact ties by catalog index and take target and competitor scores from the same scoring pass. \citet{diagnostics} asked a similar added-value question about a different component: whether features from diagnostic tests improve a selector that chooses among recommendation interfaces.

\section{Methods}

\subsection{Data and prediction task}

The paired experiment uses the October 2019 events of the REES46 cosmetics-shop dataset \citep{rees46}. We treat items viewed, added to a cart or purchased at the prediction time as targets, and we group each user's earlier events into three behavior streams: views; cart additions and purchases; and cart removals. The retrospective analysis also uses RetailRocket e-commerce interactions \citep{retailrocket}, taken from per-behavior tables in the format of the RecBole library \citep{recbole}, and the KuaiSAR-small search and recommendation dataset \citep{kuaisar}. Supplemental Table S1 and Section S2 give dataset statistics and preprocessing details.

We split each user's history chronologically. Events that share a timestamp form one group, and we assign a user's groups to training, validation and test in 70/15/15 proportions without separating simultaneous events. Users with fewer than three groups contribute only to training. Because each user is split separately, the partitions do not share a calendar cutoff. A prediction query is one user at one timestamp: its targets are the positive items at that timestamp, and its input is at most 50 earlier events. We drop queries that have no usable history. The embedding vocabulary comes from training histories and targets, the ranking catalog consists of the items that appear as training targets, and we remove test targets that are not in the catalog.

After preprocessing, REES46 has 1,761,398 training, 423,081 validation and 507,957 test queries and a ranking catalog of 40,749 items. In each epoch, every training query contributes one example: we pick one of its targets uniformly at random and sample 100 distinct negatives uniformly from the catalog, excluding all of the query's targets. Items the user interacted with earlier can still be negatives. We shuffle the 101 candidates and train with single-label cross-entropy. The sample depends only on the query, the seed and the epoch, so paired models see identical training examples. At evaluation, all of a query's targets count.

\subsection{Baseline and residual routing module}

The baseline (B) is a SASRec-style self-attentive encoder of item identities and positions. It does not use event types, which serve only to define the behavior streams. Let $h$ be its layer-normalized history representation, $e_c$ the embedding of candidate item $c$, $b_c$ an item bias and $d$ the embedding width. The baseline score is
\begin{equation}
s_B(c)=\frac{e_c^\top h}{\sqrt d}+b_c.
\label{eq:base}
\end{equation}

\begin{figure}[htbp]
\centering
\includegraphics[width=1\linewidth]{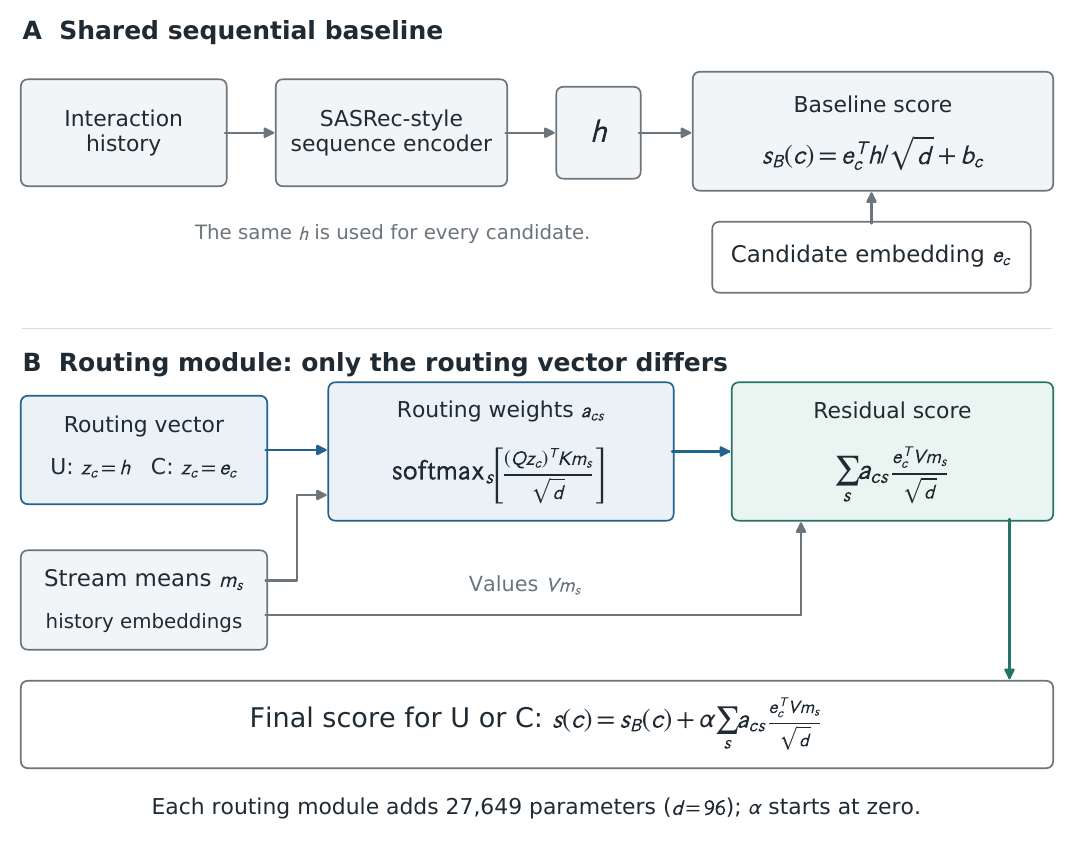}
\caption{Scoring with the residual routing module. (A) The baseline (B) scores every candidate against the same history representation $h$. (B) The two routed models use the same residual form and differ only in the routing vector that weights the behavior-stream means: the history representation for user-dependent routing (U) and the candidate embedding for candidate-dependent routing (C). In both, the residual term still depends on the candidate. The scalar $\alpha$ starts at zero, and all parameters are trained.}
\label{fig:architecture}
\end{figure}

The routing module adds an attention-weighted term to this score (Fig.~\ref{fig:architecture}). For each behavior stream $s$ that contains history events, let $m_s$ be the mean of its item embeddings, with repeated items counted once per occurrence, and let $z_c$ be the routing vector that sets the stream weights for candidate $c$.

\begin{samepage}
With learned projections $Q,K,V\in\mathbb{R}^{d\times d}$ that have no bias terms, the stream weights and the final score are
\begin{align}
a_{cs}&=\operatorname{softmax}_{s\ \mathrm{available}}
\left(\frac{(Qz_c)^\top K m_s}{\sqrt d}\right),\label{eq:attention}\\
s(c)&=s_B(c)+\alpha\sum_s a_{cs}\frac{e_c^\top V m_s}{\sqrt d},
\label{eq:residual}
\end{align}
\end{samepage}
where the softmax runs over the non-empty streams. The scalar $\alpha$ is learned, unconstrained in sign and initialized to zero, so at initialization the routed score equals the baseline score.

User-dependent routing (U) sets $z_c=h$, so every candidate receives the same stream weights. Candidate-dependent routing (C) sets $z_c=e_c$, so the weights can change from one candidate to the next. In both, the residual term still depends on the candidate through $e_c^\top V m_s$; the source of the routing vector is the only architectural difference. Each module adds $3d^2+1=27{,}649$ parameters at $d=96$ and no Transformer layers, behavior embeddings or dropout, so U and C have the same number of parameters. All baseline parameters remain trainable.

\subsection{Paired training}\label{sec:training}

We trained B, U and C with each of five seeds (2026100601--2026100605), for fifteen runs in total. Within a seed, the three models start from the same baseline weights, and U and C also share the initial weights of the routing module. Initializing the routing module leaves the baseline's random-number sequence unchanged, so the dropout draws also match across the three models. We trained every model from scratch for exactly twelve epochs and evaluated the epoch-12 weights, with no early stopping, validation-based checkpoint selection, retuning or reruns. We fixed this protocol and the evaluation queries (Section~\ref{sec:eval}) before starting the fifteen runs.

We fixed the training length because, in an earlier two-seed pilot that used early stopping on validation data, the models stopped at different epochs (Supplemental Section S3.2). Twelve epochs give every model the same number of passes over the training data.

All models use embedding width 96, two self-attention layers with four heads, histories of at most 50 events and dropout 0.1. We used AdamW \citep{adamw} with learning rate 0.001 and weight decay 0.0001, batches of 512, a 4,096-query shuffle buffer and gradient-norm clipping at 5.0; one epoch is 3,441 updates. Training and scoring use 32-bit floating point (float32), with automatic mixed precision and TensorFloat-32 disabled. Each run used one NVIDIA RTX A6000 GPU on one of three Linux hosts; checks recorded on the hosts before training show Python 3.12.13 and PyTorch 2.12.1+cu130, built for CUDA 13.0. Supplemental Section S3.6 gives further environment details, run times and the full training-loss histories.

\subsection{Evaluation queries and metrics}\label{sec:eval}

Before training, we divided the REES46 test queries into three strata according to whether their targets appear in the query's history window: all-repeat (every target appears), all-novel (none does) and mixed. The full test partition contains 243,463 all-repeat, 258,770 all-novel and 5,724 mixed queries. From each stratum we selected 1,536 queries by sorting the SHA-256 hashes of their identifiers (with a fixed prefix) and keeping the first 1,536, for 4,608 evaluation queries in total. Each selected query is weighted by its stratum's size divided by 1,536, so weighted means reproduce the stratum proportions of the 507,957-query test partition. These evaluation queries are a newly fixed subset of a test partition we had used in earlier experiments.

We evaluated each of the fifteen models once, after all training had finished. For every query, the evaluator scores all 40,749 catalog items in float32 and ranks each target against the other items using scores from that single pass. For query $q$ with catalog $\mathcal I$ and item scores $s_q(i)$, the rank of item $i$ is
\begin{equation}
r_q(i)=1+\sum_{j\in\mathcal I}\mathbf{1}
\left\{s_q(j)>s_q(i)\ \mathrm{or}\
[s_q(j)=s_q(i)\ \mathrm{and}\ j<i]\right\},
\label{eq:rank}
\end{equation}
where $\mathbf{1}\{\cdot\}$ equals one when its condition holds and zero otherwise. Exact ties are broken in favor of the item with the lower catalog index, and we add no tolerance.

Our primary metric is NDCG@10 with binary relevance, which rewards placing relevant items near the top of the list. For a query with target set $T_q$ and cutoff $K$,
\begin{equation}
g_K(q)=\frac{\sum_{i\in T_q}\mathbf{1}\{r_q(i)\le K\}/\log_2(1+r_q(i))}
{\sum_{j=1}^{\min(K,|T_q|)}1/\log_2(1+j)},
\label{eq:ndcg}
\end{equation}
and a model's NDCG@10 is the weighted mean $\widehat\mu=\sum_q w_q g_{10}(q)/\sum_q w_q$ over the evaluation queries, with $w_q$ the query weight defined above. We also report three secondary metrics at cutoffs 10 and 20, with the same weights: recall, the fraction of a query's targets ranked within the cutoff; hit rate, whether any target is ranked within the cutoff; and mean reciprocal rank (MRR), the mean of the reciprocal rank of the first target, counted as zero when no target is within the cutoff. Recall measures what fraction of the targets reach the list and hit rate whether at least one does; MRR measures how early the first one appears, and NDCG@10 reflects both how many targets appear and how high they rank.

\subsection{Statistical analysis}\label{sec:stats}

For seed $s$, the relative effect of candidate-dependent routing is $d_s=\widehat\mu_{C,s}/\widehat\mu_{B,s}-1$, and our estimate is the mean $\bar d$ of the five paired effects, which differs from the ratio of the averaged model means. The primary analysis, fixed in advance, asks whether the one-sided 95\% upper confidence bound
\begin{equation}
U_{.95}=\bar d+t_{4,.95}\frac{s_d}{\sqrt5},
\qquad t_{4,.95}=2.132,
\label{eq:upper}
\end{equation}
falls below 0.01, that is, whether a mean relative gain of 1\% or more can be ruled out. Here $s_d$ is the sample standard deviation of the five effects and $t_{4,.95}$ is the 95th percentile of Student's $t$ distribution with four degrees of freedom. The 1\% margin is a study-defined threshold rather than a measure of business value; at a baseline NDCG@10 near 0.345, it corresponds to about 0.00345. We also report a two-sided 95\% interval, using $t_{4,.975}=2.776$. Both intervals assume that the five seed effects are independent and approximately normal, which five observations cannot verify, and both are conditional on the fixed evaluation queries.

All other comparisons are exploratory: C against U, U against B, the secondary metrics and every comparison within the strata.

\subsection{Follow-up analyses}\label{sec:followup}

After seeing the primary results, we analyzed the saved ranks in three further ways. First, we recomputed all four metrics at cutoffs 1, 2, 3, 5, 10, 20, 50 and 100. Second, we split NDCG@10 into the contributions of individual rank positions, keeping each query's normalizer and weight, so that the contributions add up to the difference between the two model means rather than to the mean relative effect. Third, we classified each query in each seed by whether B, C, both or neither placed a target in the top ten; for queries that both models hit, we compared the rank of the first target.

We also rescored all fifteen trained models in float32 on a CPU (an arm64 machine running macOS) to check whether tie handling or numerical precision could explain the results. On these new scores, we compared our tie-breaking rule with five alternative orderings of exactly tied items: ascending and descending item index, a hashed order that ignores the labels, input order and uniform random order. Finally, to test whether C's losses on novel targets arise only because familiar items crowd them out, we scored a modified task in which every item in the user's recorded history is removed from the candidate list and only targets outside that history count. This task uses the 3,072 evaluation queries with at least one such target, with weights renormalized over that subset. The original GPU evaluation remains the basis of all primary results; Supplemental Section S6 describes these follow-up analyses in full.

\subsection{Retrospective analysis of Transformer models}\label{sec:retro-methods}

The retrospective analysis covers three Transformer variants on each of three datasets, with five seeds per variant. Each behavior stream is encoded by its own causal Transformer, in which a history position cannot attend to later positions. Additional attention blocks then jointly update a routing token and the stream states. The token starts from either a summary of the history or the candidate embedding. Under a local mask, streams cannot attend to one another directly, although information can pass between them through the routing token; a full mask allows direct cross-stream attention. The routing comparison changes the token's starting point from the history summary to the candidate embedding under the local mask, and the cross-stream comparison switches the candidate-dependent model from the local to the full mask. Unlike the residual module in Equations~\ref{eq:attention}--\ref{eq:residual}, which only reweights stream means, these blocks let the routing token and the stream states update each other.

Within each dataset, the three variants share an evaluation set and catalog: 4,287 queries from 3,474 users for KuaiSAR, 4,608 queries from 4,330 users for REES46 (a different set from the one in the paired experiment) and 4,176 queries from 3,935 users for RetailRocket. An earlier internal analysis plan specified twelve comparisons; we report the six Transformer comparisons whose paired query records are complete (Supplemental Section S4.5 and Table S13).

These models were selected with an earlier evaluator that computed a target's score and the catalog's scores in separate floating-point calls, so rounding could make a target count as its own competitor or change near ties. We rescored the selected checkpoints with the corrected evaluator (Equation~\ref{eq:rank}), but the checkpoint selection itself was not redone. For REES46, the original checkpoints of the two local-mask variants were no longer available, so they were retrained with a configuration chosen after inspecting validation results (up to eight epochs, with validation-based checkpoint selection). The REES46 routing comparison therefore uses two retrained models, and its cross-stream comparison pairs a retrained model with an original one; all other models use their original checkpoints. Unlike the paired experiment, these models share seed labels and training samples but are not known to share initial weights. Supplemental Section S4.2 gives the full selection history.

For each comparison, the effect is the ratio of the two models' seed-averaged means, minus one. We formed 95\% percentile intervals from 20,000 bootstrap replicates that resample both seeds and users \citep{owen}. Each replicate draws five seed labels with replacement, keeping the two models paired, and one sample of users with replacement; it keeps all of a sampled user's queries together and uses the same users for every sampled seed. The intervals are exploratory and not adjusted for multiple comparisons. Supplemental Table S12 compares them with the intervals from the earlier analysis.

\section{Results}

\subsection{Candidate-dependent routing barely changes NDCG@10}

\begin{figure}[htbp]
\centering
\includegraphics[width=1\linewidth]{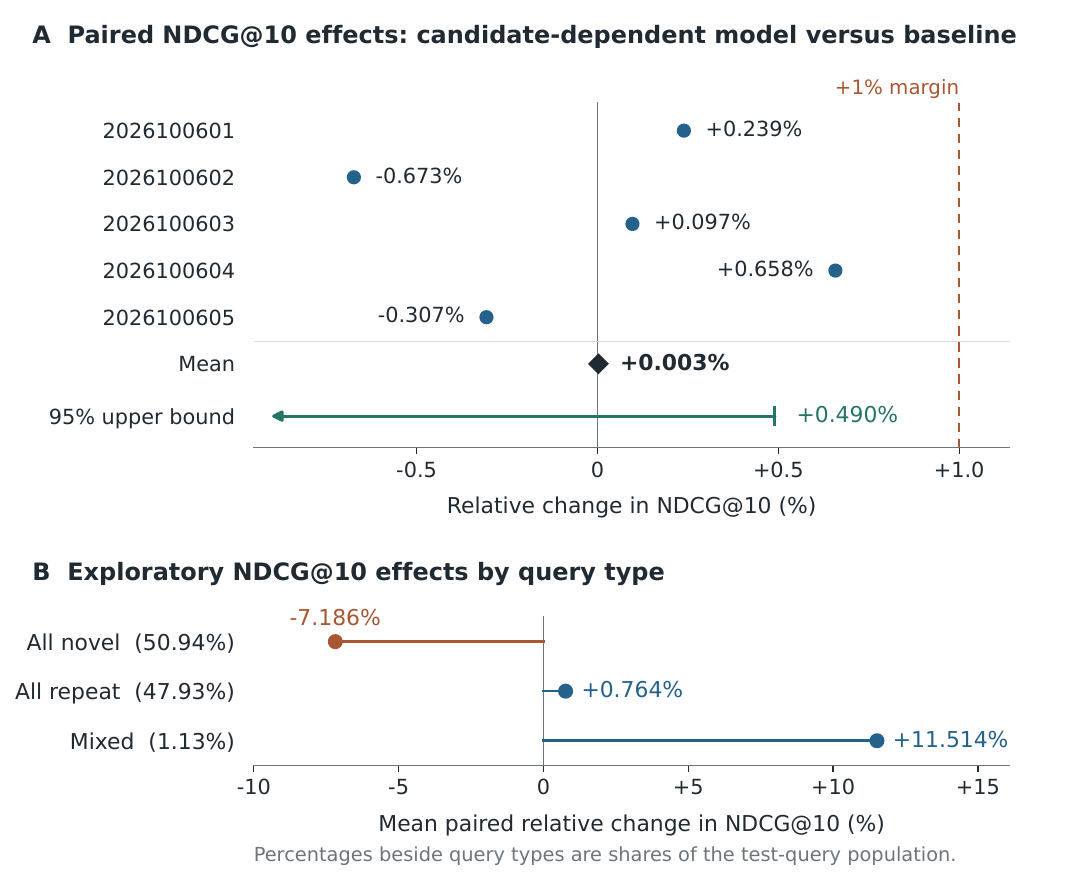}
\caption{Effect of candidate-dependent routing on NDCG@10, overall and by query type. (A) Relative change for the candidate-dependent model (C) against the baseline (B) in each of the five seeds, with their mean. The one-sided 95\% confidence bound (+0.49\%) lies below the +1\% margin set in advance. (B) Mean paired relative change in each stratum (exploratory). A query is all-novel if none of its targets appears in its history (at most 50 earlier events), all-repeat if all of them do, and mixed otherwise. Percentages beside the stratum names are shares of the test population; each stratum contributes 1,536 evaluated queries. Lines run from zero to each estimate and are not confidence intervals.}
\label{fig:paired}
\end{figure}

C and B reach almost the same mean NDCG@10, 0.344598 and 0.344592 (Table~\ref{tab:seeds}). The mean paired change is +0.003\%, and its one-sided 95\% upper bound of +0.49\% lies below the +1\% margin (Fig.~\ref{fig:paired}A). The two-sided 95\% interval is [$-0.63$\%, +0.64\%]; the five seed effects have a standard deviation of 0.512 percentage points, and three of them favor C. Under the assumptions in Section~\ref{sec:stats}, the bound excludes a mean gain of 1\% or more; because the test is one-sided, it does not show that the two models are equivalent.

\begin{table}[htbp]
\centering\small
\caption{Normalized discounted cumulative gain at cutoff ten (NDCG@10) for each of the five paired seeds at epoch 12. B, baseline; U, user-dependent routing; C, candidate-dependent routing. Values are query means weighted to the stratum proportions of the test partition. Relative changes are computed before rounding.}
\label{tab:seeds}
\begin{tabular}{lrrrr}
\toprule
Seed & B & U & C & C/B $-1$ (\%)\\
\midrule
2026100601 & 0.343049 & 0.344798 & 0.343869 & +0.2391\\
2026100602 & 0.344962 & 0.343568 & 0.342639 & -0.6734\\
2026100603 & 0.346149 & 0.346769 & 0.346484 & +0.0968\\
2026100604 & 0.343179 & 0.345101 & 0.345436 & +0.6578\\
2026100605 & 0.345625 & 0.347483 & 0.344564 & -0.3068\\
\bottomrule
\end{tabular}
\end{table}

The user-dependent model does slightly better than both. U improves on B by 0.28\% on average and C falls 0.27\% below U, with four of five seeds favoring U in each comparison (Supplemental Table S2).

\subsection{More relevant items reach the top ten, but the first one ranks lower}

Behind the flat NDCG@10, the secondary metrics move in opposite directions (Table~\ref{tab:metrics}). At cutoff ten, C raises recall by 1.806\% and hit rate by 1.701\% on average and lowers MRR by 1.085\%; all five seeds show these directions. U shows the same pattern on a smaller scale, with recall up 1.17\% and MRR down 0.32\%.

\begin{table}[htbp]
\centering\small
\caption{Ranking metrics at cutoff ten. B, baseline; U, user-dependent routing; C, candidate-dependent routing; NDCG, normalized discounted cumulative gain; MRR, mean reciprocal rank. Model means average the five seed-level weighted means, and relative changes average the five paired changes. NDCG@10 is the primary metric; the others are secondary. Supplemental Tables S3 and S4 give seed-level results at cutoffs 10 and 20.}
\label{tab:metrics}
\setlength{\tabcolsep}{4pt}
\begin{tabular}{lrrrrc}
\toprule
Metric @10 & B & U & C & C/B $-1$ (\%) & Seeds favoring C\\
\midrule
NDCG & 0.344592 & 0.345544 & 0.344598 & +0.003 & 3/5\\
Recall & 0.429810 & 0.434808 & 0.437542 & +1.806 & 5/5\\
Hit rate & 0.449208 & 0.454164 & 0.456819 & +1.701 & 5/5\\
MRR & 0.329954 & 0.328900 & 0.326372 & -1.085 & 0/5\\
\bottomrule
\end{tabular}
\end{table}

The pattern depends on list length. Recall and hit rate improve in every seed at cutoffs 3, 5, 10 and 20, but both fall on average at cutoffs 1 and 2 (Supplemental Table S14). C gets more relevant items into a short list, but not into its first one or two positions.

Comparing the two models query by query shows how these effects cancel (Supplemental Table S16). Relative to B, C gains a top-ten hit on 3.66\% of weighted query--seed pairs and loses one on 2.90\%. Most new hits arrive low in the list: in 60.10\% of them, the first target is at positions six to ten. Queries hit only by C add 0.014213 to the difference in mean NDCG@10, and queries hit only by B subtract 0.010923. On queries that both models hit, C ranks the first target lower more often than higher, and these queries subtract a further 0.003284, almost cancelling the net gain from new hits.

Splitting NDCG@10 by rank position shows the same cancellation (Table~\ref{tab:position}). C loses 0.009761 at rank one and gains 0.005186 at rank two, 0.003489 at rank three and 0.001092 across ranks four to ten, for a net difference of +0.000006. In short, C redistributes relevant items within the top ten: more targets make the list, but fewer of them come first.

\begin{table}[htbp]
\centering\small
\setlength{\tabcolsep}{4pt}
\caption{Contribution of each rank position to NDCG@10 (exploratory). Each entry is the weighted mean over queries of the discounted gain from a target at that position, divided by the query's ideal discounted cumulative gain, then averaged over the five seeds (absolute NDCG scale). The columns sum to the model means, so the C $-$ B column sums to the difference in means, not to the mean paired relative change.}\label{tab:position}
\begin{tabular}{lrrr}
\toprule
Position & Baseline B & Candidate C & C $-$ B\\
\midrule
1 & 0.266793 & 0.257033 & -0.009761\\
2 & 0.029430 & 0.034616 & +0.005186\\
3 & 0.014428 & 0.017917 & +0.003489\\
4--10 & 0.033942 & 0.035034 & +0.001092\\
Total & 0.344592 & 0.344598 & +0.000006\\
\bottomrule
\end{tabular}
\end{table}

\subsection{Repeat queries improve and novel queries get worse}

The effect of C depends on whether a query's targets appear in its history window (Fig.~\ref{fig:paired}B). On all-novel queries, NDCG@10 falls from 0.070436 for B to 0.065344 for C (mean paired change $-7.186$\%). On all-repeat queries it rises from 0.639349 to 0.644228 (+0.764\%), and on mixed queries from 0.201544 to 0.224753 (+11.514\%). Each direction holds in all five seeds.

Recall@10 splits the same way: $-4.55$\% on all-novel, +3.29\% on all-repeat and +17.92\% on mixed queries, and within each stratum recall, hit rate and NDCG change in the same direction in every seed (Supplemental Table S15). Because mixed queries contain both kinds of target, their gain could come from either.

All-novel and all-repeat queries make up 50.94\% and 47.93\% of the test population, and mixed queries only 1.13\%. After weighting, the gains on repeat and mixed queries almost exactly offset the loss on all-novel queries: their contributions to the difference in mean NDCG@10 are +0.002338, +0.000262 and $-0.002594$, which sum to +0.000006. The overall effect cannot be recovered by averaging the stratum percentages, because the weighted metric means are combined before the relative changes are computed. Supplemental Table S6 gives all stratum comparisons.

\subsection{Checks on tie handling and history items}

On the newly computed CPU scores, every tie-breaking rule we tested gives identical metrics at cutoff ten. These scores also lead to the same primary conclusion: a mean NDCG@10 change of +0.0086\% with an upper bound of +0.5267\%, below +1\% (Supplemental Table S17). CPU and GPU values of NDCG@10 differ for 188 of the 69,120 query--model--seed combinations. Because the original GPU score vectors were not saved, we cannot fully explain these differences or check the ties in the original scores.

Removing history items from the candidate list does not reveal an advantage on novel targets. In that modified task, C changes Recall@10 by $-0.26$\%, hit rate by $-0.20$\% and NDCG@10 by $-3.14$\% relative to B, and its NDCG remains below B's in all five seeds at cutoffs 10, 20, 50 and 100 (Supplemental Table S18).

\subsection{Earlier Transformer models show dataset-dependent effects}

\begin{figure}[htbp]
\centering
\includegraphics[width=1\linewidth]{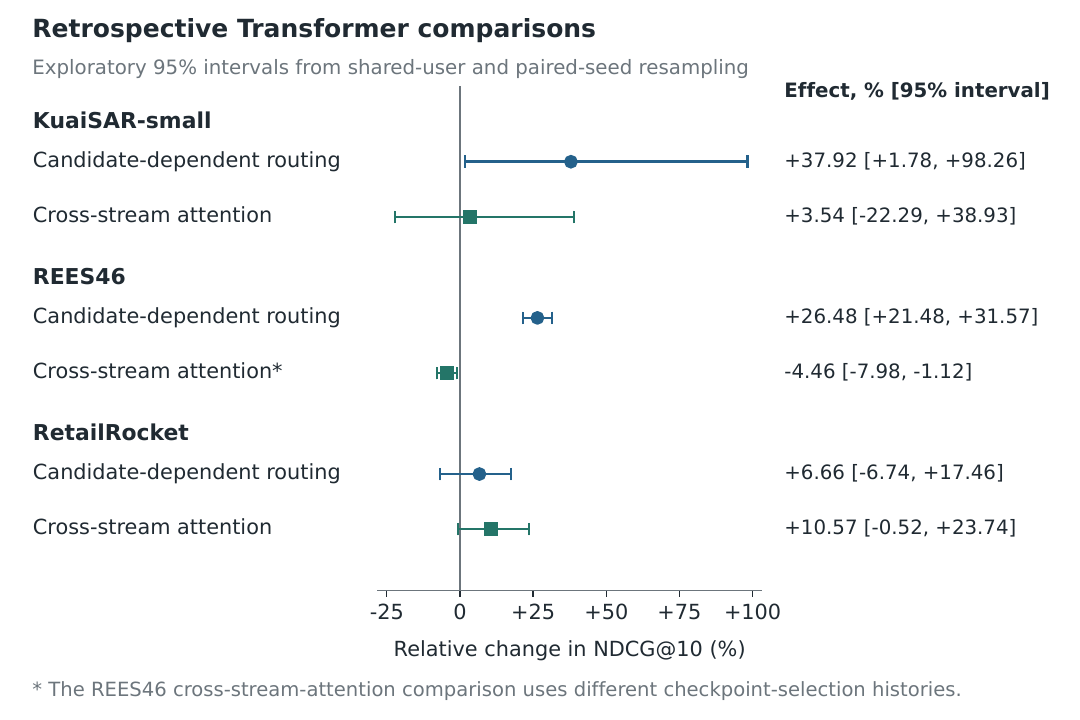}
\caption{Candidate-dependent routing and cross-stream attention in the earlier Transformer models. Points are relative changes in NDCG@10 between seed-averaged model means; bars are exploratory 95\% bootstrap intervals that resample users and paired seeds. Candidate-dependent routing starts the routing token from the candidate embedding instead of a history summary, with streams unable to attend to one another directly. Cross-stream attention lets the streams attend to one another in the candidate-dependent model. On REES46, both models in the routing comparison were retrained, and the cross-stream comparison pairs a retrained model with an original one.}
\label{fig:retrospective}
\end{figure}

In the earlier Transformer models, candidate-dependent routing improves on user-dependent routing in point estimate on all three datasets, but the size and uncertainty of the gain vary widely (Fig.~\ref{fig:retrospective}): +37.92\% on KuaiSAR (95\% interval +1.78\% to +98.26\%), +26.48\% on REES46 (+21.48\% to +31.57\%) and +6.66\% on RetailRocket ($-6.74$\% to +17.46\%). The large KuaiSAR percentage starts from a very small base: mean NDCG@10 rises from 0.001587 to 0.002188, an absolute gain of about 0.0006. The REES46 comparison involves the two retrained models described in Section~\ref{sec:retro-methods}.

Allowing the streams to attend to one another changes NDCG@10 by +3.54\% on KuaiSAR ($-22.29$\% to +38.93\%), $-4.46$\% on REES46 ($-7.98$\% to $-1.12$\%) and +10.57\% on RetailRocket ($-0.52$\% to +23.74\%). The negative REES46 result compares a retrained model with an original one, so it cannot isolate the effect of cross-stream attention. The earlier plan's rule labels an interval positive if it lies entirely above +1\%, negative if entirely below $-1$\%, equivalent if entirely within $\pm1$\%, and unresolved otherwise; by that rule the six comparisons give two positive, one negative and three unresolved results. Supplemental Tables S11 and S12 give the model means and compare these intervals with the earlier ones. With only three datasets, we do not pool the effects.

\section{Discussion}

\subsection{What the paired comparison shows}

Letting each candidate set its own stream weights barely changed average NDCG@10 in this experiment. With the same number of parameters, the user-dependent module did slightly better than the candidate-dependent one, although neither moved NDCG@10 by more than a fraction of a percent relative to the baseline.

The result concerns one specific way of using behavior information. The routing module works with the mean item embedding of each stream, which discards the order of events within the stream (U can still draw on order through the baseline's history representation when it sets its weights), and it adds no sequence encoder of its own. The experiment asks whether this extra summarize-and-weight step helps a shared sequential model. It does not measure the value of behavior-aware encoders in general, of richer candidate-dependent attention blocks, or of the published architectures discussed in Related Work.

\subsection{Recall, early ranking and repeated items}

Although average NDCG@10 barely moves, candidate-dependent routing changes what the model recommends. More queries get a relevant item into the top ten, most often near the bottom. Among queries that both models already served, most keep the same first-hit rank, but downward moves are more frequent and more costly than upward ones. The recall gains hold for lists of three to twenty items and come from queries with at least one target in the recent history. Candidate-dependent routing is thus better at filling a short list with relevant items when repeat interactions are part of the task, but it is not a better ranker overall.

The form of the residual term (Equation~\ref{eq:residual}) suggests one explanation. It compares the candidate's embedding with projected summaries of items already in the history, so the trained models may favor candidates similar to recently seen items. We have not tested this mechanism: the decompositions show where the metric changes, not which learned component causes the change. The history-exclusion task does rule out one simple version of the explanation. Even with history items removed from the candidate list, C ranks novel targets lower than B does, so crowding by familiar items cannot fully account for its losses on novel queries.

A retrieval stage that passes a short list to a reranker might value the extra relevant items, but we did not evaluate a two-stage system. RepeatNet's explicit choice between repeating and exploring \citep{repeatnet} suggests a more targeted design, in which routing also accounts for how likely the user is to repeat an item. Testing such a model would require a new time period and a stated downstream objective.

\subsection{Relation to the Transformer results}

The +26.48\% REES46 estimate from the earlier Transformer models compares candidate-dependent with user-dependent routing. The matching comparison in the paired experiment is C against U, which gave $-0.27$\%, not the +0.003\% change relative to the baseline. The two experiments also differ in architecture, evaluation queries and model selection. In the Transformers, the candidate embedding initializes a routing token that the stream states attend to, so the candidate can change the history representation itself rather than only the weights on fixed stream means. Those models were also selected through a different process, including retrained models on REES46. We cannot attribute the gap to any one of these differences; doing so would require comparing the architectures and selection procedures under the same task and training design.

\subsection{Limitations}

The paired experiment uses one dataset, one baseline implementation, one routing architecture, one training configuration and five seeds. Its confidence bound reflects only variation across training seeds on one fixed set of queries, drawn from a test partition we had used before, so it does not account for variation across queries, users or time. Training loss was still falling at epoch 12 in all fifteen runs (Supplemental Tables S8 and S9), and longer training could change the results or favor one model over another. The recall--MRR trade-off and the repeat--novel split came from exploratory analyses and need confirmation on new data.

The retrospective analysis covers only the six of twelve planned comparisons with complete query records. Its checkpoints were selected with a flawed evaluator, two of its REES46 models were retrained with a configuration chosen after inspecting validation results, and with five seeds and a stratified evaluation set its bootstrap intervals carry no general coverage guarantee. It describes these particular models and cannot support a general claim that one architecture is better.

Both experiments split each user's history separately rather than at a shared date. A stronger test would use a later time period with a common cutoff and give every model the same tuning budget.

\section{Conclusions}

On the REES46 dataset, adding candidate-dependent routing to a self-attentive sequential recommender barely changed NDCG@10: the mean change was +0.003\%, and the conditional one-sided 95\% upper bound of +0.49\% lies below the +1\% margin set in advance. The flat average hides a trade-off. The candidate-dependent model retrieves more relevant items into its top ten but, on average, ranks the first one lower, and its gains come from queries with at least one target in the recent history, while queries whose targets are all new to that history get worse. Candidate-dependent routing may therefore suit a retrieval stage judged by how many relevant items reach a short list, but that use needs to be tested against an explicit short-list objective, on a later time period, rather than assumed. In the earlier Transformer models, candidate-dependent routing beat user-dependent routing by margins that varied across datasets, but differences in architecture and model selection prevent a direct comparison with the paired experiment.

\section*{Acknowledgments}

OpenAI Codex was used under the authors' direction to support research planning, implementation, experiment execution, evaluation, verification, analysis, interpretation and manuscript preparation. The recorded historical model identifiers were \texttt{gpt-6-astra}, \texttt{gpt-5.6-sol}, \texttt{gpt-6-sol} and \texttt{gpt-6.1-sol}. Anthropic Claude (Claude Opus 5.5) was used to identify weaknesses, suggest revisions and rewrite the manuscript text and submission materials for clarity. Google Gemini 3.1 Pro provided editorial feedback on an earlier manuscript draft concerning clarity, terminology and repetitive qualifications. OpenAI ChatGPT (GPT-6 Astra Pro) assisted with rewriting the manuscript and preparing figures and submission materials to improve clarity and consistency across the report. The authors are responsible for the study and its content, including references; original and revised drafts are retained. No artificial intelligence tool is an author.

\section*{Funding}

This work was self-funded. The authors received no external grants or compute credits.

\section*{Competing Interests}

The authors declare there are no competing interests.

\section*{Author Contributions}

Han Chen and Yingrui Li contributed equally to developing the research ideas, designing the experiments, interpreting the results and critically revising the manuscript's intellectual content.

\section*{Ethics and Permissions}

The study analyzed existing, publicly available interaction datasets. No participants were recruited and no new interaction data were collected. Sources and dataset terms are described in Supplemental Information.

\section*{Data and Code Availability}

The Supplemental Information is included in this preprint. The separate supporting archives supplied with the journal submission contain saved query ranks, model and analysis code, configurations, environment records, and follow-up diagnostics. The journal submission also includes implementation-access code and a reviewer README. These archives are not included in this arXiv deposit; access enquiries should be directed to the corresponding author. Supplemental Sections S5 and S6 document their contents, extraction instructions, and reproduction commands. They support reproducing reported summaries from saved outputs, not complete reconstruction of model training or historical scoring. Original raw datasets are available from the cited providers. The supporting software and data retain their separate terms, including CC BY-NC-SA 4.0 for KuaiSAR-derived records; no dataset is relicensed by this preprint.

\FloatBarrier
\phantomsection
\addcontentsline{toc}{section}{References}
\bibliographystyle{plainnat}
\bibliography{references}

\begin{thebibliography}{20}
\providecommand{\natexlab}[1]{#1}
\providecommand{\url}[1]{\texttt{#1}}
\expandafter\ifx\csname urlstyle\endcsname\relax
  \providecommand{\doi}[1]{doi: #1}\else
  \providecommand{\doi}{doi: \begingroup \urlstyle{rm}\Url}\fi

\bibitem[Chen and Li(2026)]{diagnostics}
H.~Chen and Y.~Li.
\newblock Do evidence-reading diagnostics improve interface selection in small
  {LLM} recommenders?, 2026.
\newblock URL \url{https://arxiv.org/abs/2609.37472v1}.

\bibitem[Ferrari~Dacrema et~al.(2019)Ferrari~Dacrema, Cremonesi, and
  Jannach]{progress}
M.~Ferrari~Dacrema, P.~Cremonesi, and D.~Jannach.
\newblock Are we really making much progress? {A} worrying analysis of recent
  neural recommendation approaches.
\newblock In \emph{Proceedings of RecSys}, pages 101--109, 2019.
\newblock \doi{10.1145/3298689.3347058}.

\bibitem[Guo et~al.(2026)Guo, Chen, Zhu, and Li]{ties}
C.~Guo, H.~Chen, Y.~Zhu, and Y.~Li.
\newblock Tie handling is part of the evaluation protocol: An order-invariance
  audit for tie-heavy recommender scores, 2026.
\newblock URL \url{https://arxiv.org/abs/2609.26977v1}.
\newblock FRAME 2026 workshop.

\bibitem[Hidasi et~al.(2016)Hidasi, Karatzoglou, Baltrunas, and Tikk]{gru}
B.~Hidasi, A.~Karatzoglou, L.~Baltrunas, and D.~Tikk.
\newblock Session-based recommendations with recurrent neural networks.
\newblock In \emph{ICLR}, 2016.
\newblock URL \url{https://arxiv.org/abs/1511.06939}.

\bibitem[Holm(1979)]{holm}
S.~Holm.
\newblock A simple sequentially rejective multiple test procedure.
\newblock \emph{Scandinavian Journal of Statistics}, 6\penalty0 (2):\penalty0
  65--70, 1979.
\newblock URL \url{https://www.jstor.org/stable/4615733}.

\bibitem[Jannach and Chen(2026)]{standards}
D.~Jannach and L.~Chen.
\newblock Improving methodological standards in recommender systems offline
  evaluation.
\newblock \emph{ACM Transactions on Recommender Systems}, 4\penalty0
  (3):\penalty0 1--15, 2026.
\newblock \doi{10.1145/3800587}.
\newblock Editorial; Article 34e.

\bibitem[Kang and McAuley(2018)]{sasrec}
W.~C. Kang and J.~McAuley.
\newblock Self-attentive sequential recommendation.
\newblock In \emph{Proceedings of ICDM}, pages 197--206, 2018.
\newblock \doi{10.1109/ICDM.2018.00035}.

\bibitem[Kechinov(2020)]{rees46}
M.~Kechinov.
\newblock {eCommerce} events history in cosmetics shop.
\newblock Kaggle dataset, 2020.
\newblock URL
  \url{https://www.kaggle.com/datasets/mkechinov/ecommerce-events-history-in-cosmetics-shop}.
\newblock Version 6, 16 March 2020; October 2019 file; accessed 6 October 2026.

\bibitem[Krichene and Rendle(2020)]{sampling}
W.~Krichene and S.~Rendle.
\newblock On sampled metrics for item recommendation.
\newblock In \emph{Proceedings of KDD}, pages 1748--1757, 2020.
\newblock \doi{10.1145/3394486.3403226}.

\bibitem[Li et~al.(2019)Li, Liu, Wu, Xu, Zhao, Huang, Kang, Chen, Li, and
  Lee]{mind}
C.~Li, Z.~Liu, M.~Wu, Y.~Xu, H.~Zhao, P.~Huang, G.~Kang, Q.~Chen, W.~Li, and
  D.~L. Lee.
\newblock Multi-interest network with dynamic routing for recommendation at
  {Tmall}.
\newblock In \emph{Proceedings of CIKM}, pages 2615--2623, 2019.
\newblock \doi{10.1145/3357384.3357814}.

\bibitem[Loshchilov and Hutter(2019)]{adamw}
I.~Loshchilov and F.~Hutter.
\newblock Decoupled weight decay regularization.
\newblock In \emph{ICLR}, 2019.
\newblock URL \url{https://arxiv.org/abs/1711.05101v3}.

\bibitem[Owen(2007)]{owen}
A.~B. Owen.
\newblock The pigeonhole bootstrap.
\newblock \emph{The Annals of Applied Statistics}, 1\penalty0 (2):\penalty0
  386--411, 2007.
\newblock \doi{10.1214/07-AOAS122}.

\bibitem[Ren et~al.(2019)Ren, Chen, Li, Ren, Ma, and de~Rijke]{repeatnet}
P.~Ren, Z.~Chen, J.~Li, Z.~Ren, J.~Ma, and M.~de~Rijke.
\newblock {RepeatNet}: A repeat aware neural recommendation machine for
  session-based recommendation.
\newblock \emph{Proceedings of the AAAI Conference on Artificial Intelligence},
  33\penalty0 (1):\penalty0 4806--4813, 2019.
\newblock \doi{10.1609/aaai.v33i01.33014806}.

\bibitem[{Retailrocket}(undated)]{retailrocket}
{Retailrocket}.
\newblock Retailrocket recommender system dataset, undated.
\newblock URL
  \url{https://www.kaggle.com/datasets/retailrocket/ecommerce-dataset}.
\newblock The atomic exports used in this study are at
  https://zenodo.org/records/10998222, DOI 10.5281/zenodo.10998222; accessed 6
  October 2026.

\bibitem[Si et~al.(2023)Si, Sun, Zhang, Xu, Zang, Song, Gai, and Wen]{sesrec}
Z.~Si, Z.~Sun, X.~Zhang, J.~Xu, X.~Zang, Y.~Song, K.~Gai, and J.~R. Wen.
\newblock When search meets recommendation: Learning disentangled search
  representation for recommendation.
\newblock In \emph{Proceedings of SIGIR}, 2023.
\newblock \doi{10.1145/3539618.3591786}.

\bibitem[Sun et~al.(2023)Sun, Si, Zang, Leng, Niu, Song, Zhang, and
  Xu]{kuaisar}
Z.~Sun, Z.~Si, X.~Zang, D.~Leng, Y.~Niu, Y.~Song, X.~Zhang, and J.~Xu.
\newblock {KuaiSAR}: A unified search and recommendation dataset.
\newblock In \emph{CIKM}, 2023.
\newblock \doi{10.1145/3583780.3615123}.
\newblock URL \url{https://kuaisar.github.io/}.

\bibitem[Yang et~al.(2022)Yang, Huang, Xia, Liang, Yu, and Li]{mbht}
Y.~Yang, C.~Huang, L.~Xia, Y.~Liang, Y.~Yu, and C.~Li.
\newblock Multi-behavior hypergraph-enhanced transformer for sequential
  recommendation.
\newblock In \emph{Proceedings of KDD}, 2022.
\newblock \doi{10.1145/3534678.3539342}.

\bibitem[Yuan et~al.(2022)Yuan, Guo, He, Guo, Liu, and Tang]{mbstr}
E.~Yuan, W.~Guo, Z.~He, H.~Guo, C.~Liu, and R.~Tang.
\newblock Multi-behavior sequential transformer recommender.
\newblock In \emph{SIGIR}, pages 1642--1652, 2022.
\newblock \doi{10.1145/3477495.3532023}.

\bibitem[Zhao et~al.(2021)Zhao, Mu, Hou, Lin, Chen, Pan, Li, Lu, Wang, Tian,
  Min, Feng, Fan, Chen, Wang, Ji, Li, Wang, and Wen]{recbole}
W.~X. Zhao, S.~Mu, Y.~Hou, Z.~Lin, Y.~Chen, X.~Pan, K.~Li, Y.~Lu, H.~Wang,
  C.~Tian, Y.~Min, Z.~Feng, X.~Fan, X.~Chen, P.~Wang, W.~Ji, Y.~Li, X.~Wang,
  and J.~R. Wen.
\newblock {RecBole}: Towards a unified, comprehensive and efficient framework
  for recommendation algorithms.
\newblock In \emph{Proceedings of CIKM}, pages 4653--4664, 2021.
\newblock \doi{10.1145/3459637.3482016}.

\bibitem[Zhou et~al.(2018)Zhou, Zhu, Song, Fan, Zhu, Ma, Yan, Jin, Li, and
  Gai]{din}
G.~Zhou, X.~Zhu, C.~Song, Y.~Fan, H.~Zhu, X.~Ma, Y.~Yan, J.~Jin, H.~Li, and
  K.~Gai.
\newblock Deep interest network for click-through rate prediction.
\newblock In \emph{Proceedings of KDD}, pages 1059--1068, 2018.
\newblock \doi{10.1145/3219819.3219823}.

\end{thebibliography}

\clearpage
\phantomsection
\addcontentsline{toc}{part}{Supplemental Information}
\begin{center}
{\LARGE\bfseries Supplemental Information}\par\vspace{0.65em}
{\large Candidate-dependent residual routing trades top-of-list accuracy for recall in sequential recommendation}\par\vspace{0.5em}
Han Chen and Yingrui Li
\end{center}
\vspace{0.5em}
\noindent References cited in this supplement are included in the preceding reference list.
\setcounter{section}{0}
\setcounter{subsection}{0}
\setcounter{table}{0}
\setcounter{equation}{0}
\renewcommand{\thesection}{S\arabic{section}}
\renewcommand{\thetable}{S\arabic{table}}
\renewcommand{\theequation}{S\arabic{equation}}
\renewcommand{\theHsection}{supp.\arabic{section}}
\renewcommand{\theHsubsection}{supp.\arabic{section}.\arabic{subsection}}
\renewcommand{\theHtable}{supp.\arabic{table}}
\renewcommand{\theHequation}{supp.\arabic{equation}}
\renewcommand{\arraystretch}{1.12}
\FloatBarrier
\section{Scope and model labels}

This supplement gives the data construction, complete numerical results, model configurations and analysis history for the two experiments reported in the article. The primary experiment compares a baseline sequential recommender with two residual routing modules. Each routing module adds a correction to the baseline score by weighting history grouped by event type; these groups are called behavior streams. Training follows a fixed twelve-epoch protocol. The retrospective analysis examines three Transformer variants that model these streams, using previously selected saved model states (checkpoints). Their architectures, evaluation query sets (cohorts) and training histories differ, so their outcomes are reported separately.

The primary models are B (Baseline), U (User-dependent) and C (Candidate-dependent). The unchanged source records call the candidate-dependent model L; thus C in the article and this supplement maps to L in those records. The paired experiment is stored as \emph{Study B}, and the retrospective Transformer analysis as \emph{Study A}. These historical labels remain in archive names and script arguments, and the archive filenames keep the internal version labels under which their contents were frozen. Tables~S14--S18 report the follow-up ranking diagnostics described in Section~S6. Section~\ref{sec:tablemap} maps each table to the saved reporting files and explains the historical numbering.

\FloatBarrier
\section{Data construction and prediction task}\label{sec:data}

\subsection{Chronological split, histories and targets}

Events with the same user and timestamp are kept together. For users with at least three timestamp groups, the builder assigns groups chronologically to training, validation and test in 70/15/15 proportions. Users with fewer than three groups contribute only training events. This is a split within each user's history, rather than a backtest with one calendar cutoff shared by all users.

A query consists of all positive target items for a user at one timestamp. Its history contains at most 50 strictly preceding events. The embedding vocabulary, the set of items represented by learned vectors, is built from training history and target events; the ranking catalog contains only items observed as training targets. Out-of-catalog targets are removed, and queries without usable prior history are omitted. The split construction applies no additional iterative user/item minimum-frequency ($k$-core) filter. Table~\ref{tab:processed} gives the resulting query and catalog counts before evaluation subsets are selected.

\begin{table}[htbp]
\centering\small
\setlength{\tabcolsep}{3pt}
\caption{Processed query counts and ranking catalog sizes. Counts precede selection of the evaluation cohorts. The catalog contains training target items and excludes reserved tokens.}\label{tab:processed}
\begin{tabular}{lrrrr}
\toprule
Dataset & Training & Validation & Test & Catalog items\\
\midrule
REES46 & 1,761,398 & 423,081 & 507,957 & 40,749\\
RetailRocket & 787,209 & 246,408 & 299,730 & 225,101\\
KuaiSAR-small & 1,013,508 & 185,227 & 205,745 & 1,151,617\\
\bottomrule
\end{tabular}
\end{table}

The ranking catalog and embedding table have different sizes. REES46 has 41,091 embedding rows, including reserved entries, and 40,749 ranking items. Its largest catalog vocabulary index is 41,090. RetailRocket has 225,103 embedding rows and 225,101 ranking items; KuaiSAR-small has 3,580,702 embedding rows and 1,151,617 ranking items. Both experiments use the same REES46 catalog.

The routing streams group native event types. For REES46, views form the \texttt{general} stream, carts and purchases form \texttt{conversion}, and cart removals form the internally named \texttt{exposure} stream. That last label means cart-removal history, not advertising impressions. RetailRocket maps views to \texttt{general} and add-to-cart/transaction events to \texttt{conversion}. KuaiSAR maps clicks to \texttt{general}; likes, follows and forwards to \texttt{conversion}; query/search-click events to \texttt{search}; and impressions to \texttt{exposure}. Absent streams are masked. The grouped representation therefore does not preserve every native behavior as a separate stream.

\subsection{Source data and provenance}

\paragraph{REES46 cosmetics-shop events, October 2019.}
Histories use views, carts, cart removals and purchases; targets use views, carts and purchases. The recorded input is \texttt{2019-Oct.csv.gz} (87,330,634 bytes). Its split contains 2,703,946 training, 578,185 validation and 820,152 test events, totaling 4,102,283. The builder drops 394,569 training queries without history and 818 out-of-catalog test targets.

The internal analysis plan identifies this source as October cosmetics events. An older acquisition note gave the URL for a multi-category dataset. On 6 October 2026, the official Kaggle version-6 October comma-separated values (CSV) file was checked against the recorded preprocessing: it contains 4,102,283 data rows, nine columns and 482,542,278 decompressed bytes, with timestamps from 1 October 2019 00:00:00 to 31 October 2019 23:59:54 UTC. Running the unchanged builder on this file reproduced the event checksums for all three partitions, all three query checksums and all 47 compared manifest fields. The historical input basename was retained because it contributes to event identities.

Event checksums use ordered event identity, user, item, timestamp and normalized event type; query checksums use user, timestamp, target items and partition. Matching checksums show agreement on the events and prediction queries produced by the recorded preprocessing. They do not establish byte equality of historical CSV formatting or unused metadata. The original gzip checksum and the new decompressed-file checksum are both preserved in Supplemental Data S4. This source check performed no model scoring.

The dataset page credits Michael Kechinov and dates version 6 to 16 March 2020: \url{https://www.kaggle.com/datasets/mkechinov/ecommerce-events-history-in-cosmetics-shop}. The 2019 date describes the observation period; 2020 describes the dataset release. The page retains data-file copyright with the original authors and permits free use with attribution to Kaggle and REES46; it does not give a blanket redistribution license. The original acquisition date is unavailable. The raw file downloaded for the source check is excluded from the release.

\paragraph{RetailRocket.}
The inputs are the behavior-specific interaction tables from \url{https://zenodo.org/records/10998222} (DOI 10.5281/zenodo.10998222): \path{RetailrocketView.zip} (296,755,342 bytes), \path{RetailrocketAddtocart.zip} (266,833,184 bytes) and \path{RetailrocketTransaction.zip} (266,361,599 bytes). The data loader reads the first \texttt{.inter} table in each archive, assigns behavior from the archive name, and constructs identities from file, behavior and row before chronological sorting. No additional cross-file deduplication is claimed. Histories and targets use all three event types. The split contains 2,196,251 training, 252,007 validation and 307,843 test events; 1,407,580 training queries without history and 7,004 out-of-catalog test targets are dropped.

On 6 October 2026, all three public archives were streamed and hashed without extraction. Each archive's byte count and SHA-256 checksum matched the recorded value; the published MD5 checksums also matched. The Zenodo record credits MTandHJ, is dated 19 April 2024 and declares the Creative Commons Attribution license (CC BY 4.0). Its atomic tables are compatible with RecBole \citep{recbole}, but the uploader's conversion and deduplication options are not documented in the available records. The release follows the attribution and license of these verified Zenodo files. A separate Kaggle source lists Creative Commons Attribution-NonCommercial-ShareAlike (CC BY-NC-SA 4.0) and was not used for the file comparison. Earlier source reviews are dated 19--20 June 2026; the original acquisition date is unavailable.

\paragraph{KuaiSAR-small.}
The input is \texttt{KuaiSAR.zip} (488,077,651 bytes), the small release linked from \url{https://kuaisar.github.io/} and DOI 10.5281/zenodo.8181109. It is distinct from \texttt{KuaiSAR\_v2.zip}. The author page identifies the small release's observation window as 22--31 May 2023 and assigns CC BY-NC-SA 4.0. Histories include clicks, impressions, explicit feedback, query events and search clicks; targets are recommendation and search clicks. Expansion into these behavior events produces 7,979,856 training, 1,613,455 validation and 1,669,829 test events. The builder drops 7,949 training queries without history and 216,069 out-of-catalog test targets. The source and license were checked on 6 October 2026, following a historical review on 13 September 2026. The original download date is unavailable.

\subsection{Training examples and full-catalog metrics}

Both experiments use the same task sampler. For each query and epoch, it selects one of the query's positive targets uniformly, samples 100 distinct catalog items uniformly as negatives while excluding all of that query's targets, and shuffles the 101 candidates. Previously seen items remain eligible negatives when they are not current targets. Training uses cross-entropy with the sampled positive as the single target label. Query identity, random seed and epoch determine the sampled candidates and their order, so paired models receive the same examples. Evaluation retains all of the query's positive targets.

For evaluation query $x$, let $T_x$ be its positive item set, $\mathcal I$ the catalog and $s_x(i)$ the score of item $i$ from scoring the full catalog. The corrected full-catalog rank is
\begin{equation}
r_x(i)=1+\sum_{j\in\mathcal I}\mathbf{1}\{s_x(j)>s_x(i)\ \text{or}\ [s_x(j)=s_x(i)\ \text{and}\ j<i]\}.
\end{equation}
The evaluator takes the target score from the same 32-bit floating-point (float32) computation used for its competitors. Lower catalog indices break exact ties; no epsilon is used. This rule is independent of candidate input order and does not average over random tie permutations. Catalog scoring uses query batches of 16 and item chunks of 16,384.

Binary-relevance normalized discounted cumulative gain (NDCG) at cutoff $K$ is
\begin{equation}
g_K(x)=\frac{\sum_{i\in T_x}\mathbf{1}\{r_x(i)\le K\}/\log_2(1+r_x(i))}
{\sum_{j=1}^{\min(K,|T_x|)}1/\log_2(1+j)}.
\end{equation}
Each query receives weight $w_x$ equal to its test-stratum count divided by the number selected from that stratum. The NDCG@10 model summary is $\widehat\mu=\sum_x w_xg_{10}(x)/\sum_xw_x$. All other query metrics use the same weighted mean. Recall is the fraction of positives within the cutoff, and hit rate indicates whether at least one positive is retrieved. Reciprocal rank is the reciprocal of the first positive's rank if it lies within the cutoff, and zero otherwise. Mean reciprocal rank (MRR) averages these query values. 

\FloatBarrier
\section{Paired residual-routing experiment}\label{sec:paired}

\subsection{Models and initialization}

The baseline is the study's model based on Self-Attentive Sequential Recommendation (SASRec) \citep{sasrec}. It encodes item identities and positions, with no behavior-type embeddings. Event types define the history groups used by the routing module. It is not claimed to reproduce the original reference implementation exactly. With embedding width $d$, layer-normalized history representation $h$, candidate embedding $e_c$ and candidate bias $b_c$, it scores
\begin{equation}
s_B(c)=e_c^\top h/\sqrt d+b_c.
\end{equation}
For each available behavior stream $s$, the routing module computes the mean $m_s$ of its history-item embeddings, counting repeated items once per occurrence. It uses bias-free projections $Q,K,V\in\mathbb R^{d\times d}$ and a learned scalar $\alpha$. Let $z_c$ denote the routing vector and $a_{cs}$ the weight assigned to stream $s$ for candidate $c$:
\begin{align}
a_{cs}&=\operatorname{softmax}_{s\text{ available}}\big((Qz_c)^\top K m_s/\sqrt d\big),\\
s(c)&=s_B(c)+\alpha\sum_s a_{cs}\,e_c^\top V m_s/\sqrt d.
\end{align}
U uses $z_c=h$; C uses $z_c=e_c$. The routing-vector source is their sole architectural difference. The scalar $\alpha$ is initialized to zero and is unconstrained in sign. Both routing modules add $3d^2+1=27,649$ parameters at $d=96$, and all baseline parameters remain trainable. U and C have the same parameter count; B has fewer parameters. The routing module adds no Transformer layers, behavior embeddings or dropout. It weights stream means without jointly updating a routing token and stream states, unlike the retrospective architecture.

Seeds 2026100601--2026100605 each initialize B, U and C from the same baseline weights. U and C additionally share routing module initialization. The initialization procedure preserves the baseline random-number stream, including subsequent dropout draws. Equality of the initial weights is verified for this experiment; matching seed labels alone do not establish it for the retrospective models.

\subsection{Training protocol and preceding pilot}

All fifteen final fits train from scratch for exactly twelve epochs and use epoch 12 for evaluation. The protocol was fixed before launch. The final experiment uses no validation-based training adjustment, early stopping, outcome-based tuning, warm start or retry. The baseline has width 96, four attention heads, two sequence layers, maximum history 50 and dropout 0.1. AdamW, Adam with decoupled weight decay \citep{adamw}, uses learning rate 0.001 and weight decay 0.0001. Batch size is 512, with 100 negatives, a 4,096-query shuffle buffer and gradient-norm clipping at 5.0. Every epoch contains 1,761,398 queries and 3,441 updates. Computation uses float32, with automatic mixed precision and TensorFloat-32 disabled.

A preceding two-seed pilot used training and validation data, validation-based checkpoint selection and early stopping. Its six fits selected epochs 8, 8, 8, 4, 4 and 7 and completed 8, 8, 8, 6, 6 and 8 epochs, respectively. These unequal durations motivated using the same final epoch for every model. The pilot's instructions and script guards specify training/validation use; they do not provide an independent reconstruction of the original private execution history. Twelve epochs equalize the number of training passes in the final experiment. They do not establish convergence, optimal training duration or equal wall-clock cost.

\subsection{Fixed evaluation cohort}

Before final training, stored REES46 test queries were ordered by deterministic hash values, and 1,536 were selected from each of three strata (query subgroups), for 4,608 queries overall. An all-repeat query has every positive item in its retained history of at most 50 strictly prior events; an all-novel query has none there, and a mixed query has some. An item seen only before that window is novel under this definition. These strata do not measure lifetime novelty or item cold start.

The processed test population contains 258,770 all-novel, 243,463 all-repeat and 5,724 mixed queries. Weights are the respective population counts divided by 1,536, giving weighted denominator 507,957. These weights reproduce the test partition's stratum proportions; deterministic hash selection does not itself establish probability-sampling guarantees. The saved field \texttt{inclusion\_probability} stores the stratum sample fraction, and historical files call the weighting inverse-probability weighting. All-query estimates retain the recorded weights. The cohort was newly fixed, but it is a subset of a previously used test partition, not a new time window or wholly unseen test partition. The original evaluation scored each final model once after training and preservation checks, against all 40,749 ranking items. The later CPU diagnostic in Section~S6 is separate from that evaluation.

\subsection{Primary statistic and secondary outcomes}

For seed $s$, the primary effect is $d_s=\widehat\mu_{C,s}/\widehat\mu_{B,s}-1$. The effect of interest is the mean of five paired relative effects, $\bar d$, rather than a ratio of pooled model means. With sample standard deviation $s_d$, the one-sided 95\% upper confidence bound and descriptive two-sided 95\% confidence interval are
\begin{equation}
U_{.95}=\bar d+t_{4,.95}\,s_d/\sqrt5,\qquad
C_{.95}=\bar d\pm t_{4,.975}\,s_d/\sqrt5,
\end{equation}
Here $t_{4,.95}=2.132$ and $t_{4,.975}=2.776$ are quantiles of Student's $t$ distribution with four degrees of freedom. The prespecified decision asks whether the upper bound is below a +1\% relative gain. This is a study-defined threshold, not a calibrated measure of business utility; near baseline NDCG@10 of 0.345 it corresponds to about 0.00345 absolute NDCG.

The mean paired C/B effect is +0.002698\%, its one-sided upper bound is +0.490404\%, and its descriptive two-sided interval is [$-0.632472$\%, $+0.637869$\%]. The sample standard deviation of seed effects is 0.511548 percentage points. The bound is below the +1\% criterion. Its uncertainty concerns training-seed variation conditional on this fixed cohort. The $t$ interval assumes independent, approximately normally distributed seed effects, an assumption that is difficult to assess from five observations. It does not include uncertainty from selecting queries or supply a population-coverage guarantee. The analysis is not an equivalence test.

Table~\ref{tab:paired-contrasts} gives all paired NDCG differences. Tables~\ref{tab:metrics-10} and~\ref{tab:metrics-20} report every model/seed outcome at cutoffs 10 and 20. C/U, U/B and other metrics are descriptive. Relative effects are calculated before rounding; dividing displayed model means need not reproduce a mean paired ratio.

\begin{table}[htbp]
\centering\footnotesize
\setlength{\tabcolsep}{3pt}
\caption{Paired NDCG@10 contrasts for all five seeds. Absolute differences use the NDCG scale; relative differences are percentages. C/B is the primary contrast; C/U and U/B are descriptive.}\label{tab:paired-contrasts}
\begin{tabular}{lrrrrr}
\toprule
Seed & C $-$ B & C $-$ U & U $-$ B & C/U $-1$ (\%) & U/B $-1$ (\%)\\
\midrule
2026100601 & +0.00082011 & -0.00092968 & +0.00174979 & -0.2696 & +0.5101\\
2026100602 & -0.00232286 & -0.00092863 & -0.00139423 & -0.2703 & -0.4042\\
2026100603 & +0.00033500 & -0.00028591 & +0.00062090 & -0.0824 & +0.1794\\
2026100604 & +0.00225743 & +0.00033504 & +0.00192240 & +0.0971 & +0.5602\\
2026100605 & -0.00106033 & -0.00291821 & +0.00185788 & -0.8398 & +0.5375\\
\bottomrule
\end{tabular}
\end{table}

\begin{table}[htbp]
\centering\small
\setlength{\tabcolsep}{3pt}
\caption{Weighted query metrics at cutoff 10 for every epoch-12 checkpoint in the paired experiment. B denotes the baseline, U user-dependent routing and C candidate-dependent routing.}\label{tab:metrics-10}
\begin{tabular}{lrrrr}
\toprule
Model / seed & NDCG@10 & Recall@10 & MRR@10 & Hit rate@10\\
\midrule
B / 2026100601 & 0.343049 & 0.425816 & 0.329085 & 0.445334\\
B / 2026100602 & 0.344962 & 0.429973 & 0.330582 & 0.449318\\
B / 2026100603 & 0.346149 & 0.434908 & 0.330473 & 0.454692\\
B / 2026100604 & 0.343179 & 0.426366 & 0.328928 & 0.445636\\
B / 2026100605 & 0.345625 & 0.431984 & 0.330703 & 0.451058\\
U / 2026100601 & 0.344798 & 0.438126 & 0.327147 & 0.457117\\
U / 2026100602 & 0.343568 & 0.429795 & 0.328268 & 0.449385\\
U / 2026100603 & 0.346769 & 0.437372 & 0.329697 & 0.457468\\
U / 2026100604 & 0.345101 & 0.431822 & 0.328925 & 0.451146\\
U / 2026100605 & 0.347483 & 0.436925 & 0.330461 & 0.455704\\
C / 2026100601 & 0.343869 & 0.438917 & 0.325175 & 0.458224\\
C / 2026100602 & 0.342639 & 0.432226 & 0.325817 & 0.451856\\
C / 2026100603 & 0.346484 & 0.438078 & 0.328338 & 0.457093\\
C / 2026100604 & 0.345436 & 0.439582 & 0.326677 & 0.458968\\
C / 2026100605 & 0.344564 & 0.438905 & 0.325855 & 0.457953\\
\bottomrule
\end{tabular}
\end{table}

\begin{table}[htbp]
\centering\small
\setlength{\tabcolsep}{3pt}
\caption{Weighted query metrics at cutoff 20 for every epoch-12 checkpoint. These outcomes are descriptive secondary measures.}\label{tab:metrics-20}
\begin{tabular}{lrrrr}
\toprule
Model / seed & NDCG@20 & Recall@20 & MRR@20 & Hit rate@20\\
\midrule
B / 2026100601 & 0.359525 & 0.490730 & 0.333617 & 0.509691\\
B / 2026100602 & 0.359072 & 0.486571 & 0.334577 & 0.507937\\
B / 2026100603 & 0.360277 & 0.491965 & 0.334308 & 0.511388\\
B / 2026100604 & 0.358354 & 0.487403 & 0.333085 & 0.506862\\
B / 2026100605 & 0.359533 & 0.488003 & 0.334495 & 0.507147\\
U / 2026100601 & 0.357951 & 0.490254 & 0.330795 & 0.509644\\
U / 2026100602 & 0.358697 & 0.489693 & 0.332505 & 0.509870\\
U / 2026100603 & 0.361794 & 0.497387 & 0.333829 & 0.517473\\
U / 2026100604 & 0.359519 & 0.489354 & 0.332963 & 0.508872\\
U / 2026100605 & 0.361551 & 0.493140 & 0.334405 & 0.512228\\
C / 2026100601 & 0.358794 & 0.498062 & 0.329301 & 0.517269\\
C / 2026100602 & 0.358551 & 0.495394 & 0.330209 & 0.515239\\
C / 2026100603 & 0.360837 & 0.495611 & 0.332274 & 0.514824\\
C / 2026100604 & 0.360054 & 0.498133 & 0.330724 & 0.518458\\
C / 2026100605 & 0.358873 & 0.495782 & 0.329842 & 0.515224\\
\bottomrule
\end{tabular}
\end{table}

\subsection{Exploratory ranking and repeat-stratum summaries}

After evaluation was complete, the saved query outputs were summarized by metric and repeat stratum. Table~\ref{tab:secondary-contrasts} reports overall contrasts at cutoff 10, and Table~\ref{tab:stratum-contrasts} gives all three NDCG@10 contrasts within each stratum. The accompanying machine-readable outputs contain all three contrasts and all eight metrics, both overall and within strata.

C has higher recall and hit rate than B at cutoff 10 in all five seeds, but lower MRR in every seed. Its NDCG@10 effects are negative for all-novel queries and positive for all-repeat and mixed queries in every seed. These summaries describe different aspects of the same ranked outputs. They do not establish why the changes occur or provide additional confirmatory confidence claims.

\begin{table}[htbp]
\centering\small
\setlength{\tabcolsep}{3pt}
\caption{Overall descriptive contrasts at cutoff 10. Relative effects are the mean of five paired seed ratios, not ratios of the model means. A positive seed favors the numerator model.}\label{tab:secondary-contrasts}
\begin{tabular}{llrrr}
\toprule
Contrast & Metric & Mean difference & Relative (\%) & Positive seeds\\
\midrule
C/B & NDCG@10 & +0.000006 & +0.003 & 3/5\\
C/B & Recall@10 & +0.007732 & +1.806 & 5/5\\
C/B & Hit rate@10 & +0.007611 & +1.701 & 5/5\\
C/B & MRR@10 & -0.003582 & -1.085 & 0/5\\
C/U & NDCG@10 & -0.000945 & -0.273 & 1/5\\
C/U & Recall@10 & +0.002734 & +0.632 & 5/5\\
C/U & Hit rate@10 & +0.002655 & +0.587 & 4/5\\
C/U & MRR@10 & -0.002527 & -0.768 & 0/5\\
U/B & NDCG@10 & +0.000951 & +0.277 & 4/5\\
U/B & Recall@10 & +0.004998 & +1.168 & 4/5\\
U/B & Hit rate@10 & +0.004956 & +1.108 & 5/5\\
U/B & MRR@10 & -0.001054 & -0.319 & 0/5\\
\bottomrule
\end{tabular}
\end{table}

\begin{table}[htbp]
\centering\small
\setlength{\tabcolsep}{3pt}
\caption{Exploratory NDCG@10 contrasts within repeat strata. Novelty is defined against the retained history of at most 50 strictly prior events. Relative effects average the five paired seed ratios. Positive seeds favor the first model in each contrast.}\label{tab:stratum-contrasts}
\begin{tabular}{llrrr}
\toprule
Stratum & Contrast & Mean difference & Relative (\%) & Positive seeds\\
\midrule
all-novel & C/B & -0.005092 & -7.186 & 0/5\\
all-novel & C/U & -0.002650 & -3.792 & 0/5\\
all-novel & U/B & -0.002442 & -3.430 & 1/5\\
all-repeat & C/B & +0.004879 & +0.764 & 5/5\\
all-repeat & C/U & +0.000517 & +0.082 & 3/5\\
all-repeat & U/B & +0.004362 & +0.682 & 5/5\\
mixed & C/B & +0.023209 & +11.514 & 5/5\\
mixed & C/U & +0.013893 & +6.587 & 5/5\\
mixed & U/B & +0.009316 & +4.623 & 5/5\\
\bottomrule
\end{tabular}
\end{table}

\subsection{Execution environment, runtime and training history}

The paired runs used three Linux hosts, with one NVIDIA RTX A6000 graphics processing unit (GPU) per run. Checks recorded on 5 October 2026, before final training, identify Python 3.12.13, PyTorch 2.12.1+cu130 built for CUDA 13.0 and NumPy 2.5.0 on all three hosts. These observations concern the assigned interpreters before training, not a new version check for every fit. An additional record identifies cuDNN package 9.20.0.48 on one host; the corresponding versions for the other hosts are unavailable. Supplemental Data S5 contains the recovered records and their source checksums. Torch intra- and inter-op thread counts were one. Two persistent central processing unit (CPU) workers prepared the data, with four prefetched batches. The specified batch order and CUDA update order were preserved. The compatibility field \texttt{workspace\_layers=2} is retained in the configuration but adds no routing-token--stream attention blocks to the routing module. The specific operating-system distribution, point release and NVIDIA driver are not recorded for these hosts. This environment description applies to the paired experiment. The available records do not establish the hardware and imported software versions for every retained retrospective Transformer checkpoint.

Table~\ref{tab:runtime} gives observed training wall times, evaluation times and peak allocated GPU tensor memory. Training wall time includes startup and checkpoint writes; evaluation time comes from the evaluator. Memory is in mebibytes (MiB; $2^{20}$ bytes) and does not include all memory occupied on the device. Differences in scheduling and device contention prevent attributing timing differences solely to the models. Tables~\ref{tab:loss-early} and~\ref{tab:loss-late} give all epoch-mean training losses. Loss decreases from epoch 11 to epoch 12 in every fit.

\begin{table}[htbp]
\centering\small
\setlength{\tabcolsep}{3pt}
\caption{Observed training and evaluation costs. Peak memory is allocated GPU tensor memory. Scheduling and device contention differed among runs, so timing differences cannot be attributed solely to the models.}\label{tab:runtime}
\begin{tabular}{lrrrr}
\toprule
Model / seed & Train (min) & Train peak (MiB) & Evaluate (s) & Eval. peak (MiB)\\
\midrule
B / 2026100601 & 55.37 & 540.4 & 31.74 & 130.1\\
B / 2026100602 & 50.52 & 540.4 & 28.23 & 130.1\\
B / 2026100603 & 67.86 & 540.4 & 33.44 & 130.1\\
B / 2026100604 & 54.84 & 540.4 & 32.04 & 130.1\\
B / 2026100605 & 47.05 & 540.4 & 29.22 & 130.1\\
U / 2026100601 & 52.21 & 551.3 & 37.63 & 140.2\\
U / 2026100602 & 69.55 & 551.3 & 35.29 & 140.2\\
U / 2026100603 & 58.14 & 551.3 & 33.87 & 140.2\\
U / 2026100604 & 53.61 & 551.3 & 29.85 & 140.2\\
U / 2026100605 & 72.69 & 551.3 & 38.71 & 140.2\\
C / 2026100601 & 70.52 & 576.4 & 34.91 & 230.2\\
C / 2026100602 & 58.22 & 576.4 & 31.85 & 230.2\\
C / 2026100603 & 53.91 & 576.4 & 35.68 & 230.2\\
C / 2026100604 & 71.63 & 576.4 & 38.77 & 230.2\\
C / 2026100605 & 57.92 & 576.4 & 33.92 & 230.2\\
\bottomrule
\end{tabular}
\end{table}

\begin{table}[htbp]
\centering\footnotesize
\setlength{\tabcolsep}{3pt}
\caption{Mean training losses for epochs 1--6 of all fifteen fits. Each epoch contains 1,761,398 queries and 3,441 optimizer updates.}\label{tab:loss-early}
\begin{tabular}{lrrrrrr}
\toprule
Model / seed & Epoch 1 & Epoch 2 & Epoch 3 & Epoch 4 & Epoch 5 & Epoch 6\\
\midrule
B / 2026100601 & 3.1383 & 2.2913 & 1.7979 & 1.5140 & 1.3450 & 1.2320\\
B / 2026100602 & 3.1300 & 2.2575 & 1.7734 & 1.5014 & 1.3376 & 1.2266\\
B / 2026100603 & 3.1225 & 2.2634 & 1.7902 & 1.5112 & 1.3412 & 1.2268\\
B / 2026100604 & 3.1219 & 2.2698 & 1.7850 & 1.5087 & 1.3407 & 1.2276\\
B / 2026100605 & 3.1250 & 2.2803 & 1.7856 & 1.5053 & 1.3380 & 1.2268\\
U / 2026100601 & 3.1150 & 2.2797 & 1.7758 & 1.4922 & 1.3263 & 1.2168\\
U / 2026100602 & 3.1080 & 2.2488 & 1.7536 & 1.4815 & 1.3201 & 1.2124\\
U / 2026100603 & 3.0998 & 2.2522 & 1.7694 & 1.4916 & 1.3248 & 1.2128\\
U / 2026100604 & 3.1009 & 2.2602 & 1.7630 & 1.4870 & 1.3223 & 1.2121\\
U / 2026100605 & 3.1012 & 2.2716 & 1.7666 & 1.4859 & 1.3206 & 1.2115\\
C / 2026100601 & 3.1034 & 2.2727 & 1.7715 & 1.4861 & 1.3185 & 1.2084\\
C / 2026100602 & 3.0962 & 2.2425 & 1.7481 & 1.4745 & 1.3121 & 1.2033\\
C / 2026100603 & 3.0890 & 2.2477 & 1.7641 & 1.4841 & 1.3161 & 1.2034\\
C / 2026100604 & 3.0888 & 2.2522 & 1.7563 & 1.4789 & 1.3131 & 1.2031\\
C / 2026100605 & 3.0899 & 2.2622 & 1.7581 & 1.4773 & 1.3117 & 1.2028\\
\bottomrule
\end{tabular}
\end{table}

\begin{table}[htbp]
\centering\footnotesize
\setlength{\tabcolsep}{3pt}
\caption{Mean training losses for epochs 7--12 of all fifteen fits. The final checkpoint was specified before the experiment. Training loss alone does not show that ranking performance has stopped improving.}\label{tab:loss-late}
\begin{tabular}{lrrrrrr}
\toprule
Model / seed & Epoch 7 & Epoch 8 & Epoch 9 & Epoch 10 & Epoch 11 & Epoch 12\\
\midrule
B / 2026100601 & 1.1491 & 1.0839 & 1.0306 & 0.9834 & 0.9428 & 0.9080\\
B / 2026100602 & 1.1458 & 1.0817 & 1.0281 & 0.9835 & 0.9428 & 0.9077\\
B / 2026100603 & 1.1439 & 1.0777 & 1.0242 & 0.9789 & 0.9379 & 0.9039\\
B / 2026100604 & 1.1455 & 1.0794 & 1.0270 & 0.9810 & 0.9402 & 0.9060\\
B / 2026100605 & 1.1441 & 1.0794 & 1.0258 & 0.9802 & 0.9411 & 0.9062\\
U / 2026100601 & 1.1361 & 1.0728 & 1.0208 & 0.9751 & 0.9355 & 0.9012\\
U / 2026100602 & 1.1337 & 1.0711 & 1.0190 & 0.9756 & 0.9363 & 0.9020\\
U / 2026100603 & 1.1315 & 1.0671 & 1.0148 & 0.9704 & 0.9315 & 0.8973\\
U / 2026100604 & 1.1322 & 1.0681 & 1.0171 & 0.9724 & 0.9330 & 0.8998\\
U / 2026100605 & 1.1306 & 1.0683 & 1.0160 & 0.9715 & 0.9333 & 0.8993\\
C / 2026100601 & 1.1272 & 1.0644 & 1.0128 & 0.9676 & 0.9287 & 0.8944\\
C / 2026100602 & 1.1248 & 1.0627 & 1.0111 & 0.9680 & 0.9287 & 0.8948\\
C / 2026100603 & 1.1222 & 1.0581 & 1.0062 & 0.9625 & 0.9236 & 0.8904\\
C / 2026100604 & 1.1231 & 1.0593 & 1.0087 & 0.9644 & 0.9254 & 0.8927\\
C / 2026100605 & 1.1216 & 1.0602 & 1.0080 & 0.9642 & 0.9261 & 0.8920\\
\bottomrule
\end{tabular}
\end{table}

\FloatBarrier
\section{Retrospective routed-Transformer analysis}\label{sec:retrospective}

\subsection{Architecture and training configurations}

Each behavior stream is encoded separately by a causal Transformer. Its representation $m_s$ is the encoded state of its final valid event. User-dependent routing starts from the mean $u$ of the available stream states; candidate-dependent routing starts from candidate embedding $e_c$. The saved model names are \texttt{user-routed-streams}, \texttt{candidate-local-routing} and \texttt{candidate-full-routing}.

Let $z$ be the initial routing token ($u$ or $e_c$), and $\mathcal S$ the available streams. Attention blocks jointly update the routing token and stream tokens to $z',m'_1,\ldots,m'_S$. A local mask permits each stream to attend to itself and the shared routing token. A full mask also permits direct attention between different streams. Multiple local blocks can still transfer information indirectly through the routing token, so local routing does not completely isolate streams.

With a learned scalar-valued function $R$ and vector concatenation $[\cdot,\cdot]$, the route weights and combined representation are
\begin{equation}
a_{cs}=\operatorname{softmax}_{s\in\mathcal S}R([z,m'_s]),\qquad
f_c=z'+\sum_{s\in\mathcal S}a_{cs}m'_s.
\end{equation}
The function $R$ uses the initial routing token $z$, while the combined representation includes the updated token $z'$. A learned vector-valued function $O$ and layer normalization, denoted $\operatorname{LN}$, give score $e_c^\top\operatorname{LN}(O([f_c,e_c]))/\sqrt d+b_c$. The routing contrast changes the initial routing token in the local model. The cross-stream-attention contrast enables direct stream links while retaining candidate-dependent routing.

Table~\ref{tab:transformer-recipes} gives the training configurations actually used. The historical training and selection procedure allowed up to eight epochs, selected the checkpoint with highest validation NDCG and stopped after two epochs without improvement. Training batch size was 512, with 100 negatives, a 4,096-query shuffle buffer and maximum history 50. Twelve configurations were planned per method/dataset; Transformer selection used mean validation NDCG across five stored variants. Tuning and checkpoint selection used the historical evaluator before the rank correction. Per-model records supply actual selected epochs, training histories, parameters and links to source files. Default settings are not used to fill gaps in the recorded configurations.

\begin{table}[htbp]
\centering\small
\setlength{\tabcolsep}{3pt}
\caption{Training configurations used in the retrospective Transformer analysis, taken from the saved input summaries. Seq. layers denotes the number of causal sequence-encoder layers. Each model uses two query--stream attention blocks and weight decay 0.0001.}\label{tab:transformer-recipes}
\begin{tabular}{llrrrrr}
\toprule
Dataset & Family & Width & Heads & Seq. layers & Dropout & Learning rate\\
\midrule
RetailRocket & Transformer & 64 & 2 & 1 & 0.0 & 0.001\\
REES46 & Transformer & 64 & 2 & 1 & 0.0 & 0.001\\
KuaiSAR-small & Transformer & 64 & 2 & 2 & 0.1 & 0.0003\\
\bottomrule
\end{tabular}
\end{table}

\subsection{Checkpoint selection and rank correction}

All five REES46 user-dependent local and candidate-dependent local checkpoints are replacements for unavailable originals. Both models in the REES46 routing contrast therefore use replacements. Its cross-stream-attention contrast compares replacement local checkpoints with historical full-routing checkpoints. The replacement training configuration was selected after validation inspection and then fixed across five seeds. Its associated contract targeted the two hypotheses with the strongest results in validation selection, including REES46 Transformer routing. The original replacement protocol declared no use of the test partition before that freeze, but the inspected values and earlier access history were not preserved. These records cannot establish that earlier test outcomes were unseen. Replacement weights are selected later models, not replicas of the missing weights. The other reported models use historical checkpoints.

The historical evaluator computed a target score in a narrow scoring call and compared it with scores from a separate computation over the full catalog. Floating-point rounding that depended on the scoring-call shape could make the target count itself as a higher-scoring competitor or alter a near tie. The corrected evaluator uses the target's score from the same float32 computation used for all catalog items, as defined in Section~\ref{sec:data}. The reported ranks come from rescoring the selected checkpoints with that evaluator. Historical validation and selection were not rerun. Matched before/after query-level test records for the correction are unavailable, so its empirical rank-change count and effect on the reported estimates cannot be quantified.

\subsection{Cohorts and model means}

The retrospective cohorts contain 4,287 queries from 3,474 users for KuaiSAR-small, 4,608 from 4,330 users for REES46, and 4,176 from 3,935 users for RetailRocket. Models within each dataset share its cohort and catalog. The analysis uses 45 complete corrected query-rank files: three models, three datasets and five seed labels, totaling 196,065 query rows. Shared seed labels and training samples support paired observations but do not establish identical initial tensors.

Table~\ref{tab:transformer-means} reports the nine model means. Each averages five seed-specific weighted query means. These cohorts and model-selection histories differ from the paired residual experiment, so an apparent difference between their effects cannot identify an architectural, training or evaluation cause.

\begin{table}[htbp]
\centering\small
\setlength{\tabcolsep}{3pt}
\caption{Nine model means supporting the six retrospective Transformer contrasts. Each entry averages five seed-specific weighted query means. REES46 user-dependent and candidate-dependent local models use replacement checkpoints.}\label{tab:transformer-means}
\begin{tabular}{llrr}
\toprule
Dataset & Model & Seeds & NDCG@10\\
\midrule
KuaiSAR-small & Candidate-dependent, full & 5 & 0.002266\\
KuaiSAR-small & Candidate-dependent, local & 5 & 0.002188\\
KuaiSAR-small & User-dependent, local & 5 & 0.001587\\
REES46 & Candidate-dependent, full & 5 & 0.298720\\
REES46 & Candidate-dependent, local & 5 & 0.312651\\
REES46 & User-dependent, local & 5 & 0.247203\\
RetailRocket & Candidate-dependent, full & 5 & 0.271816\\
RetailRocket & Candidate-dependent, local & 5 & 0.245838\\
RetailRocket & User-dependent, local & 5 & 0.230485\\
\bottomrule
\end{tabular}
\end{table}

\subsection{Relative effects and exploratory intervals}

The retrospective relative effect is a ratio of equally weighted seed-level model means, $\bar\mu_{\mathrm{treat}}/\bar\mu_{\mathrm{ctrl}}-1$, where the subscripts identify the treatment and control models in each contrast. It differs from the paired experiment's mean of within-seed ratios. The historical bootstrap sampled users independently for each seed draw, formed an absolute-difference interval and divided its endpoints by the fixed observed control mean. This can treat repeated users across seeds as additional user information and leaves denominator uncertainty fixed.

The revised calculation resamples five paired seed labels and samples users once per replicate. Each sampled user contributes all of that user's weighted queries, and the same user draw is used for every selected seed. The calculation then recomputes both model means and their ratio. The 2.5th and 97.5th percentiles of 20,000 replicates form the exploratory interval. Deterministic user and seed random states are recorded. Using the same sampled users across seeds retains their shared contribution, consistent with crossed resampling \citep{owen}; it does not establish confidence-interval coverage with only five seeds. The ordinary user bootstrap also does not reproduce the cohort's original stratified sampling design, although inclusion weights are retained.

Table~\ref{tab:transformer-intervals} shows both interval sets for every complete contrast. The revised calculation reproduces the historical point estimates. Before revision, the preserved historical procedure reproduced all six absolute bootstrap endpoints to $10^{-12}$. Numerical checks covered an independently written small reference, duplicated seed labels with identical user effects, zero effects, doubled treatment values and varying weighted denominators. The corrected duplicated-seed draws were unchanged to tolerance; the historical interval width fell to about 46\% of its one-seed width. These checks validate the calculation, not confidence-interval coverage.

\begin{table}[htbp]
\centering\small
\setlength{\tabcolsep}{3pt}
\caption{Historical and revised exploratory 95\% intervals for all six complete Transformer contrasts. Historical intervals resampled users independently for each seed draw and fixed the ratio denominator; revised intervals use the same sampled users across seeds and recompute both model means.}\label{tab:transformer-intervals}
\begin{tabular}{llrr}
\toprule
Dataset & Contrast & Historical interval (\%) & Revised interval (\%)\\
\midrule
KuaiSAR-small & Conditioning & [+8.99, +66.07] & [+1.78, +98.26]\\
KuaiSAR-small & Cross-stream attention & [-19.43, +27.64] & [-22.29, +38.93]\\
REES46 & Conditioning & [+22.63, +29.96] & [+21.48, +31.57]\\
REES46 & Cross-stream attention & [-7.83, -1.29] & [-7.98, -1.12]\\
RetailRocket & Conditioning & [-6.62, +16.49] & [-6.74, +17.46]\\
RetailRocket & Cross-stream attention & [-0.50, +20.98] & [-0.52, +23.74]\\
\bottomrule
\end{tabular}
\end{table}

The original plan defined a $\pm1\%$ region: intervals above +1\% were labeled materially positive, below $-1$\% materially negative, wholly inside the region practically equivalent, and otherwise unresolved. Those labels are descriptive and per-comparison in this reanalysis; they are not simultaneous guarantees or independently validated utility thresholds. No reported revised interval receives an equivalence label. The underlying checkpoint-selection limitations remain regardless of interval width.

\subsection{Original plan and present scope}

The internal implementation and analysis plan is dated 11 September 2026 and is supplied in Supplemental Data S4 under its original filename, \texttt{RECSYS\_F4P876\_TORS\_STUDY\_PREREGISTRATION.md}. No external registry accession is claimed. It specified twelve comparisons across the model families labeled Occurrence and Transformer in the plan. Table~\ref{tab:original-plan} gives their full disposition. This article reports the six Transformer contrasts because complete paired query ranks are available for them; the six occurrence contrasts lack those records in the release.

\begin{table}[htbp]
\centering\small
\setlength{\tabcolsep}{3pt}
\caption{Status of all twelve comparisons in the internal analysis plan dated 11 September 2026. All reported Transformer comparisons have five paired seed labels. Comparisons without complete paired query ranks are excluded. Occurrence is the original plan's name for the excluded model family.}\label{tab:original-plan}
\begin{tabular}{P{1.10in}P{.95in}P{1.55in}P{2.35in}}
\toprule
Dataset & Family & Contrast & Current evidence\\
\midrule
KuaiSAR-small & Occurrence & Conditioning & Excluded; paired ranks incomplete\\
KuaiSAR-small & Occurrence & Cross-stream attention & Excluded; paired ranks incomplete\\
KuaiSAR-small & Transformer & Conditioning & Reported; paired ranks complete\\
KuaiSAR-small & Transformer & Cross-stream attention & Reported; paired ranks complete\\
REES46 & Occurrence & Conditioning & Excluded; paired ranks incomplete\\
REES46 & Occurrence & Cross-stream attention & Excluded; paired ranks incomplete\\
REES46 & Transformer & Conditioning & Reported; paired ranks complete\\
REES46 & Transformer & Cross-stream attention & Reported; paired ranks complete\\
RetailRocket & Occurrence & Conditioning & Excluded; paired ranks incomplete\\
RetailRocket & Occurrence & Cross-stream attention & Excluded; paired ranks incomplete\\
RetailRocket & Transformer & Conditioning & Reported; paired ranks complete\\
RetailRocket & Transformer & Cross-stream attention & Reported; paired ranks complete\\
\bottomrule
\end{tabular}
\end{table}

The original plan used sign-flip tests grouped by user against zero effect, assigning each user the same sign across seeds and adjusting across all twelve comparisons with Holm's procedure \citep{holm}. Those tests are distinct from the bootstrap intervals. The current analysis does not redefine the family as six tests or make new adjusted-test claims. Earlier aggregate estimates and unadjusted $p$-values in S4 document the analysis history and are not additional findings of this article.

The planned random-effects synthesis and prediction interval are omitted because three datasets provide too little information for a reliable between-dataset interval. Sampled-candidate and broad baseline findings are omitted because complete supporting query records are unavailable in the release. The planned broader analyses of effect modifiers and computational performance are outside the present analysis of saved ranks. The paired experiment's repeat-stratum summaries are separately identified as exploratory.

\FloatBarrier
\section{Reproducibility, file contents and licenses}\label{sec:reproducibility}

The supporting archives described in this section were supplied with the journal submission and are separate from this arXiv deposit, which contains the article and Supplemental Information. Access enquiries should be directed to the corresponding author. The journal submission additionally includes Supplemental Data S7 (implementation-access code, explicit-path model-use commands, the reviewer README, and synthetic CPU verification records). S7 does not replace S1--S6. The instructions below describe the saved-output analyses once those archives have been obtained.

Supplemental Data S1, S2 and S3 each contain fifteen retrospective query-rank files and an explanatory text file, for KuaiSAR-small, REES46 and RetailRocket, respectively. S4 contains the fifteen paired-experiment query outputs and the shared source code, effective configurations, cohorts, summaries, analysis records and earlier extraction instructions. Archives S1--S4 share a common directory structure. Extract them into the same empty directory; they create an \texttt{evidence/} directory. From that directory, run the following commands, using a new output directory named \texttt{NEW\_OUTPUT}:

{\footnotesize
\begin{verbatim}
python3 study_a_repair_v5/replay_study_a_public_v12.py \
  --package-root . --out NEW_OUTPUT/study_a_replayed.json

python3 saved_analysis_v5/analyze_saved_outputs_public_v12.py \
  --study-b-extracted deposit_candidate/study_b/evaluation \
  --compare-v4 revision_analysis_v1 \
  --out NEW_OUTPUT/study_b_replayed

python3 replay_retained_reporting_v8.py --package-root . \
  --out NEW_OUTPUT/reporting_tables
\end{verbatim}
}
A trailing backslash continues a shell command onto the next line. The retrospective replay requires Python 3.10 or later and NumPy 2.x; the paired replay uses the Python standard library. Supplemental Data S5 contains the recovered environment observations and is not an input to these analysis commands. Supplemental Data S6 contains the later diagnostic subset, with separate commands in Section~S6.

The supporting archives support recomputing reported metrics and statistics from saved ranks. They do not by themselves rerun raw-data preprocessing, model training, checkpoint selection or model scoring. Third-party raw data, stored interaction histories and checkpoint files are excluded. Source checksums, data loaders and deterministic preprocessing commands are supplied for readers who obtain the source datasets under their respective terms. Historical validation summaries remain distinct from the corrected test aggregates.

The paired-experiment records include configurations, checksums of initial states, all twelve loss histories, and final and resume checkpoint metadata. The supplied scripts check query identities, cohort membership, ranks, weights and all eight metrics. Reported weighted means are reproducible to absolute tolerance $10^{-12}$ from the saved ranks.

The supporting archives retain all 112 protocol-listed Python files and all 60 query-output files unchanged. Private authorization messages and personal file paths have been removed from public metadata; separate original/public checksums preserve the link to the original records. The scientific settings and results are unchanged. The table map below supersedes the earlier title and numbering in the S4 README.

\subsection{Table-to-file map}\label{sec:tablemap}
The reporting command writes CSV files to \texttt{NEW\_OUTPUT/reporting\_tables/}. Their filenames retain the earlier numbering. Use the map below rather than inferring the current table number from a filename. Main Table 1 uses \path{Study_B_primary_per_seed.csv}; main Table 2 uses \path{Table_S8_Study_B_descriptive_overall.csv}.

\begingroup
\footnotesize
\setlength{\tabcolsep}{3pt}
\renewcommand{\arraystretch}{1.25}
\noindent\begin{tabular}{@{}P{0.09\linewidth}P{0.87\linewidth}@{}}
\toprule
Table & Saved source and relevant subset \\
\midrule
S1 & \path{Table_S10_preprocessing.csv} \\
S2 & \path{Table_S1_Study_B_paired_contrasts.csv} \\
S3 & \path{Tables_S2_S3_Study_B_secondary_metrics.csv}; cutoff 10 \\
S4 & \path{Tables_S2_S3_Study_B_secondary_metrics.csv}; cutoff 20 \\
S5 & \path{Table_S8_Study_B_descriptive_overall.csv} \\
S6 & \path{Table_S9_Study_B_descriptive_strata.csv}; NDCG@10 \\
S7 & \path{Table_S4_Study_B_runtime_memory.csv} \\
S8 & \path{Tables_S5_S6_Study_B_epoch_losses.csv}; epochs 1--6 \\
S9 & \path{Tables_S5_S6_Study_B_epoch_losses.csv}; epochs 7--12 \\
S10 & \path{Table_S11_Transformer_effective_recipes.csv} \\
S11 & \path{Table_S7_nine_routed_Transformer_means.csv} \\
S12 & \path{Table_S12_shared_user_seed_intervals.csv}; revised intervals, with earlier intervals as described below \\
S13 & \path{STUDY_A_PLAN_SCOPE_INVENTORY_v12.csv}; original-plan inventory in the extracted \texttt{evidence/} directory, not a replay output \\
\bottomrule
\end{tabular}
\endgroup

For Table S12, the earlier interval column comes from the six matching \path{*-transformer-*.json} records under \path{evidence/deposit_candidate/study_a_evidence/primary-contrasts/}. Each endpoint is \texttt{contrast.confidence\_low} or \texttt{contrast.confidence\_high}, multiplied by 100 and divided by \texttt{contrast.control\_mean}. These historical limits are shown for comparison; the revised exploratory results use the shared-user bootstrap. Table S13 records the disposition of all twelve planned comparisons without promoting excluded historical estimates to current findings.

Historical B/U/L identifiers remain unchanged in the saved files, with manuscript C corresponding to saved L.

\subsection{Licenses}
Dataset terms are retained separately from the software license. The MIT software license applies only to original software listed in the code-license manifest. KuaiSAR-derived query and target records retain CC BY-NC-SA 4.0; RetailRocket-derived records retain CC BY 4.0 from the cited atomic release. REES46 records retain provider attribution and the source copyright notice, without assigning a blanket Creative Commons license. Existing public-dataset identifiers remain to preserve pairing and deterministic user ordering. Third-party software notices and source-specific data terms accompany the relevant files.

\FloatBarrier
\section{Exploratory recall, ordering and score diagnostics}\label{sec:new-diagnostics}

These analyses followed inspection of the primary and secondary results. The first three tables below use the original GPU evaluation's saved positive ranks; the final two concern new CPU scoring of the retained epoch-12 checkpoints. Neither set changes the prespecified NDCG@10 analysis. All use the same 4,608-query cohort and original stratum-count weights unless the changed-task subset is explicitly identified.

\subsection{Recall across cutoffs and query types}

We recomputed all four metrics at cutoffs 1, 2, 3, 5, 10, 20, 50 and 100. At each cutoff, ideal DCG uses the full eligible target set, not the number retrieved. Relative effects are means of five paired seed ratios. A ratio is undefined when the denominator model has a zero metric; the machine-readable summaries mark those subgroup cases rather than replacing the five-seed estimand with a favorable subset or a ratio of pooled means. Overall effects in Table~\ref{tab:cutoff} have five defined ratios.

\begin{table}[htbp]
\centering\small
\setlength{\tabcolsep}{4pt}
\caption{Exploratory cutoff sensitivity from original historical ranks. Entries are mean paired C/B relative changes (percent). The last columns count seeds favoring C for recall and hit rate. All cutoffs use the original candidate sets, targets and stratum-count weights. NDCG@10 remains the primary outcome; no new primary cutoff is selected.}\label{tab:cutoff}
\begin{tabular}{rrrrrcc}
\toprule
$K$ & Recall & Hit rate & NDCG & MRR & Recall seeds & Hit seeds\\
\midrule
1 & -3.6655 & -3.6603 & -3.6603 & -3.6603 & 0/5 & 0/5\\
2 & -0.5986 & -0.7513 & -1.4785 & -2.0956 & 1/5 & 0/5\\
3 & +1.4463 & +1.3299 & -0.3198 & -1.2681 & 5/5 & 5/5\\
5 & +1.3752 & +1.2705 & -0.2859 & -1.2214 & 5/5 & 5/5\\
10 & +1.8062 & +1.7015 & +0.0027 & -1.0851 & 5/5 & 5/5\\
20 & +1.5688 & +1.4953 & +0.0195 & -1.0615 & 5/5 & 5/5\\
50 & +1.2751 & +1.0138 & +0.0119 & -1.0686 & 4/5 & 4/5\\
100 & +0.8834 & +0.6118 & -0.0711 & -1.0792 & 5/5 & 4/5\\
\bottomrule
\end{tabular}
\end{table}

Table~\ref{tab:coverage-strata} shows that the recall gain is concentrated in repeat and mixed queries. A mixed query can contain both recently seen and unseen targets, so its total gain does not identify a novel-target improvement. The original stratum weights are retained. Giving the three strata equal weight evaluates a different query mix, with mixed queries receiving one-third rather than 1.13\% of the total weight.

\begin{table}[htbp]
\centering\small
\setlength{\tabcolsep}{4pt}
\caption{Exploratory C/B changes by query type at cutoff ten, using original ranks. Shares refer to the complete test partition; each evaluated stratum contains 1,536 queries. Metric changes are means of five paired seed-relative effects (percent). Every seed favors C for recall, hit rate and NDCG in repeat and mixed queries; every seed favors B for those metrics in all-novel queries.}\label{tab:coverage-strata}
\begin{tabular}{lrrrrr}
\toprule
Stratum & Share (\%) & Recall & Hit rate & NDCG & MRR\\
\midrule
All-novel & 50.9433 & -4.5490 & -4.6252 & -7.1862 & -9.6983\\
All-repeat & 47.9298 & +3.2887 & +3.0563 & +0.7636 & -0.5539\\
Mixed & 1.1269 & +17.9248 & +7.5018 & +11.5140 & +2.8232\\
\bottomrule
\end{tabular}
\end{table}

\subsection{Position and paired-query decompositions}

For a given rank position, the NDCG contribution is the weighted sum of its discount whenever a positive occupies that position, divided by each query's ideal DCG and by the total query weight. We average that contribution over the five seeds. Summing positions 1--10 yields the average model NDCG@10; subtracting B from C yields +0.0000058708. Main Table~3 reports positions 1, 2, 3 and 4--10 combined. The full ten-position and per-seed data are in Supplemental Data S6.

Paired queries are classified by whether each model has a positive within the first ten positions. Table~\ref{tab:transitions} gives their shares and contributions to the absolute metric differences. C gains a hit on 3.6613\% of weighted query--seed observations and loses one on 2.9002\%. Pooling the original weights across seeds, 60.1038\% of newly gained first hits are at positions 6--10. Net contributions from newly hit and lost-hit queries add 0.00660093 to recall; changes in the number of retrieved positives on both-hit queries add 0.00113094. Worse first-hit ranks on both-hit queries offset the gains in MRR and NDCG. These are identities calculated from paired ranks, not a causal attribution to any learned score component.

\begin{table}[htbp]
\centering\small
\setlength{\tabcolsep}{4pt}
\caption{Exploratory paired-query decomposition at cutoff ten from original saved ranks. Shares and absolute metric contributions average over five seeds using original query weights. The first four rows partition all query--seed observations. The final three rows partition the both-hit row; their shares are also expressed relative to all query weight.}\label{tab:transitions}
\begin{tabular}{lrrrr}
\toprule
Paired outcome & Share (\%) & $\Delta$Recall & $\Delta$NDCG & $\Delta$MRR\\
\midrule
B only & 2.9002 & -0.028362 & -0.010923 & -0.005929\\
C only & 3.6613 & +0.034963 & +0.014213 & +0.008377\\
Both hit & 42.0206 & +0.001131 & -0.003284 & -0.006029\\
Neither hit & 51.4179 & +0.000000 & +0.000000 & +0.000000\\
\multicolumn{5}{l}{Within queries hit by both models:}\\
C first hit earlier & 5.1862 & +0.000252 & +0.008747 & +0.011180\\
Same first-hit rank & 30.5293 & +0.000766 & +0.000728 & +0.000000\\
C first hit later & 6.3051 & +0.000113 & -0.012759 & -0.017208\\
\bottomrule
\end{tabular}
\end{table}

\subsection{New CPU scoring and exact-tie sensitivity}

The audit recovered all fifteen retained checkpoints and the original query cache, checked their hashes, and generated float32 catalog scores for every original query. The recorded diagnostic environment was macOS 14.6 on arm64, Python 3.12.14, PyTorch 2.12.1 and NumPy 2.3.5, using one CPU thread, query batches of 16 and catalog chunks of 16,384. Models were in evaluation mode, without gradients, automatic mixed precision or TensorFloat-32. This environment differs from the original GPU environment described in Section~S3.6.

The independent reference compared the original comparator, ascending item index, descending item index, a label-independent hashed order, input order and uniform permutations within exact-tie blocks. A single coherent order applies to all positives in a query; expectations under uniform permutations are calculated for each metric rather than by applying a nonlinear metric to an average rank. Unequal scores are never permuted. Generated tests include exhaustive small tied rankings, cutoff-crossing ties, multiple positives and candidate-order changes.

All 69,120 query--model--seed score vectors contain some exact catalog tie, but only 84 have a positive within an exact-tie block. No exact tie affects the first position or crosses the rank-10/rank-11 boundary. All tested policies produce identical cutoff-ten NDCG, recall, hit rate and MRR. This result is about the new CPU vectors, not the unretained original GPU scores.

\begin{table}[htbp]
\centering\small
\setlength{\tabcolsep}{4pt}
\caption{Original primary result and separate CPU score diagnostic for NDCG@10. Effects and bounds are percentages and retain the same five-seed calculation. All tested exact-tie policies give identical cutoff-ten metrics on the CPU vectors. The CPU result does not replace the historical estimate or reconstruct historical GPU tie behavior.}\label{tab:cpu-check}
\begin{tabular}{lrr}
\toprule
Score source & Mean paired C/B & One-sided 95\% upper\\
\midrule
Original GPU evaluation & +0.002698 & +0.490404\\
New CPU diagnostic & +0.008627 & +0.526662\\
\bottomrule
\end{tabular}
\end{table}

Compared with the historical positive ranks, new CPU inference changes many ranks deep in the catalog and changes 188 query--model--seed NDCG@10 values. The tested CPU batch, chunk and final-reduction precision alternatives did not explain the selected historical discrepancies. The active evaluator takes the positive's score from the same full score vector as its competitors. A legacy separate-score function contains the earlier self-comparison problem but is not invoked by the active paired-study evaluation path; no justified evaluator patch was identified. Missing historical full scores prevent complete reconstruction of the original score differences.

\subsection{Recorded-history exclusion and novel targets}

The exploratory changed task retains only targets absent from the recorded history and removes all recorded-history items from every model's candidate set. It includes the 3,072 original queries with at least one novel target (1,536 all-novel and 1,536 mixed) and conditions the original weights on their combined test-partition count of 264,494. The 1,536 all-repeat queries are omitted because they have no eligible positive. Positive ranks are recomputed within the reduced candidate set; original score values and checkpoints are unchanged. The task concerns catalog items absent from a recent history window, not products absent from training or lifetime novelty.

\begin{table}[htbp]
\centering\small
\setlength{\tabcolsep}{4pt}
\caption{Exploratory changed-task diagnostic on new CPU scores. Only targets absent from the recorded history are retained, all recorded-history competitors are excluded, and 3,072 original queries with at least one novel target are evaluated. Original weights are renormalized over this subset. Entries are mean paired C/B relative changes (percent). The last column counts seeds favoring C for NDCG. This is neither a cold-start experiment nor a replacement for the original evaluation.}\label{tab:masked-novel}
\begin{tabular}{rrrrrc}
\toprule
$K$ & Recall & Hit rate & NDCG & MRR & NDCG seeds\\
\midrule
1 & -12.3233 & -12.3293 & -12.3293 & -12.3293 & 0/5\\
2 & -4.0882 & -3.9869 & -6.2650 & -7.1758 & 0/5\\
3 & -2.8691 & -2.6988 & -5.1816 & -6.2535 & 1/5\\
5 & -5.2148 & -5.1563 & -6.1217 & -6.7087 & 1/5\\
10 & -0.2634 & -0.2022 & -3.1374 & -5.0403 & 0/5\\
20 & -1.8052 & -1.7661 & -3.5464 & -5.1044 & 0/5\\
50 & -2.4266 & -2.3980 & -3.6313 & -5.0945 & 0/5\\
100 & -1.7152 & -1.7406 & -3.2299 & -5.0297 & 0/5\\
\bottomrule
\end{tabular}
\end{table}

Removing history competitors does not reveal an overall discovery advantage. Masked-novel NDCG is negative in every seed at cutoffs 10, 20, 50 and 100. A narrow mixed-novel Hit@100 effect is +0.5164\% in three seeds, while its recall and NDCG effects are negative. The complete task, subgroup and cutoff summaries preserve that exception without treating it as a general improvement.

\subsection{Files, replay commands and evidence boundary}

Supplemental Data S6 extracts to \path{ranking_diagnostics/}. Its \path{historical/} directory contains independently recomputed summaries from the original rank files in S4. The \path{cpu_rank_outputs/} directory contains derived original-task and history-excluded positive ranks for all fifteen CPU evaluations; \path{cpu_tie_metrics/} contains the corresponding per-query metrics for every tested tie policy. These are derived outcomes, not full score matrices. The original checkpoints, raw query cache and approximately 11.3 GB of complete new CPU score arrays are not embedded. Original historical GPU full score vectors were not saved.

With S1--S4 extracted as described in Section~S5 and S6 extracted beside the resulting \path{evidence/} directory, run:

{\footnotesize
\begin{verbatim}
python3 ranking_diagnostics/recompute_historical.py \
  --evaluation evidence/deposit_candidate/study_b/evaluation \
  --out NEW_OUTPUT/historical_diagnostics

python3 ranking_diagnostics/replay_cpu_diagnostics.py \
  --evaluation evidence/deposit_candidate/study_b/evaluation \
  --out NEW_OUTPUT/cpu_diagnostics
\end{verbatim}}

The first command uses the Python standard library; the second also requires NumPy. Both refuse to overwrite an existing output directory. They reproduce summaries from saved ranks and metric arrays, not model inference or the exact-tie structure of full score matrices. Independent small-ranking reference tests are supplied in \path{ranking_diagnostics/reference/}. The archive's README gives the commands and the source-to-result map below.

{\footnotesize
\setlength{\tabcolsep}{3pt}
\renewcommand{\arraystretch}{1.2}
\noindent\begin{tabular}{@{}P{0.12\linewidth}P{0.84\linewidth}@{}}
\toprule
Table & File in S6, or file produced by its replay \\
\midrule
Main 3 & \path{historical/historical_rank_contributions.csv} \\
S14 & \path{historical/historical_C_B_summary.csv}; overall, all cutoffs \\
S15 & Same file; the three strata at cutoff 10 \\
S16 & \path{historical/historical_query_transitions.csv} \\
S17 & \path{historical/REPLAY_VERIFICATION.json} and \path{cpu_replay/CPU_REPLAY_VERIFICATION.json} \\
S18 & \path{cpu_replay/cpu_C_B_summary.csv}; history-excluded, novel-containing view \\
\bottomrule
\end{tabular}}

All 552,960 stored original per-query metric values reproduce exactly from the historical ranks. The diagnostic replay also reproduces the saved CPU per-query metrics exactly and confirms 188 changed NDCG@10 values relative to the aligned historical records. These checks verify the released derived outputs; they do not substitute for unavailable historical score-level evidence.

\end{document}